\documentclass[pra,aps,twocolumn,nopacs,superscriptaddress,nofootinbib,reprint]{revtex4} %---
\usepackage{amsmath}  \usepackage{amssymb}  \usepackage{amsfonts}  \usepackage{bm}  \usepackage{bbm}   \usepackage{braket}  \usepackage{color}  \usepackage{comment}  \usepackage{dcolumn}  \usepackage{enumerate}  \usepackage{epsfig}  \usepackage{gensymb}  \usepackage{graphicx}  \usepackage{indentfirst}  \usepackage{lmodern}  \usepackage{mathrsfs}  \usepackage{mathtools}  \usepackage{psfrag}  \usepackage{pst-all}  \usepackage{soul}  \usepackage{units}  \usepackage{xcolor}
\usepackage{float} %---APS---SI---arxiv
\usepackage[colorlinks,linkcolor=blue,citecolor=blue,urlcolor=blue,hyperindex,driverfallback=dvipdfm]{hyperref}  \usepackage[T1]{fontenc} %---APS---ACS---SI---arxiv

\def\ii{{\rm i}}  \def\ee{{\rm e}}

        \def\Eb{{\bf E}}      \def\fb{{\bf f}}    \def\gb{{\bf g}}  \def\Hb{{\bf H}}          \def\kb{{\bf k}}      \def\Pb{{\bf P}}  \def\pb{{\bf p}}  \def\Qb{{\bf Q}}    \def\Rb{{\bf R}}  \def\rb{{\bf r}}  \def\Sb{{\bf S}}     %--- bold vectors
\def\xx{\hat{\bf x}}  \def\yy{\hat{\bf y}}  \def\zz{\hat{\bf z}}    \def\rr{\hat{\bf r}}    \def\eh{\hat{\bf e}}  \def\RR{\hat{\bf R}}  
\def\eps{\epsilon}  \def\epsh{\epsilon_{\rm h}}  \def\epsc{\epsilon_{\rm 1}}
\def\Gg{\mathcal{G}}  \def\kh{{k_{\rm h}}}  \def\Qh{{Q_{\rm h}}}  \def\tQh{{\tilde{Q}_{\rm h}}}

\begin{document} %---APS---SI---arxiv
%\renewcommand{\thefigure}{S\arabic{figure}} %---SI
%\renewcommand{\theequation}{S\arabic{equation}} %---SI
%\renewcommand{\thetable}{S\arabic{table}} %---SI
%\renewcommand{\thesection}{S\arabic{section}} %---SI
%\renewcommand{\thepage}{S\arabic{page}} %---SI
%\def\bibsection{\section*{\refname}} %---SI---arxiv

% =========================================================
% --- title, affiliations, abstract -----------------------
% =========================================================
\title{Direct generation of entangled photon pairs in nonlinear optical waveguides}

% --- APS affiliations ------------------------------------
\author{\'Alvaro~Rodr\'{\i}guez~Echarri}
\affiliation{ICFO-Institut de Ciencies Fotoniques, The Barcelona Institute of Science and Technology, 08860 Castelldefels (Barcelona), Spain}

\author{Joel~D.~Cox}
\affiliation{Center for Nano Optics, University of Southern Denmark, Campusvej 55, DK-5230 Odense M, Denmark}
\affiliation{Danish Institute for Advanced Study, University of Southern Denmark, Campusvej 55, DK-5230 Odense M, Denmark}

\author{F.~Javier~Garc\'{\i}a~de~Abajo}
\email{javier.garciadeabajo@nanophotonics.es}
\affiliation{ICFO-Institut de Ciencies Fotoniques, The Barcelona Institute of Science and Technology, 08860 Castelldefels (Barcelona), Spain}
\affiliation{ICREA-Instituci\'o Catalana de Recerca i Estudis Avan\c{c}ats, Passeig Llu\'{\i}s Companys 23, 08010 Barcelona, Spain}

% --- abstract --------------------------------------------
\begin{abstract}
Entangled photons are pivotal elements in emerging quantum information technologies. While several schemes are available for the production of entangled photons, they typically require the assistance of cumbersome optical elements to couple them to other components involved in logic operations. Here, we introduce a scheme by which entangled photon pairs are directly generated as guided mode states in optical waveguides. The scheme relies on the intrinsic nonlinearity of the waveguide material, circumventing the use of bulky optical components. Specifically, we consider an optical fiber under normal illumination, so that photon down-conversion can take place to waveguide states emitted with opposite momentum into a spectral region populated by only two accessible modes. By additionally configuring the external illumination to interfere different incident directions, we can produce maximally entangled photon-pair states, directly generated as waveguide modes with conversion efficiencies that are competitive with respect to existing macroscopic schemes. These results should find application in the design of more efficient and compact quantum optics devices.
\end{abstract}

\maketitle

% =========================================================
% --- introduction ----------------------------------------
% =========================================================
\section{Introduction}

As quantum information processing is reaching a mature state, different platforms that materialize quantum entanglement are being intensely explored \cite{ZG00,RBH01,LHB01,M02_1}. Among them, the generation of entangled photon pairs via nonlinear light-matter interactions is highly appealing for practical implementation, where photons---being capable of traversing enormous distances at the ultimate speed while interacting weakly with their environment---are ideal carriers of information \cite{IMT13,LCL17}. In this context, the intrinsically weak interaction of light with matter is both a blessing and a curse, in that propagating photons are less sensitive to decoherence, but are difficult to manipulate because they cannot be easily brought to interact \cite{CVL14}. Efficient harvesting of generated entangled photon pairs in optical device architectures presents further technological challenges that impede development of all-optical quantum information networks.
%\cite{G1971}
 
Quantum entanglement has traditionally been encoded in the polarization (or spin angular momentum) state of photons funnelled into the weakly guided modes supported by optical fibers \cite{LVS05,LNW20}. Alternatively, the orbital angular momentum (OAM) state of light constitutes an infinite basis set in which photon entanglement is accessed by twisting the light wavefront \cite{MVW01,FLP12,MEH16}. Recently, optical metasurfaces capable of generating light in arbitrary spin and OAM states have been employed to produce well-collimated streams of entangled photons \cite{SFM18,SAK21}.

Entangled photon pairs are typically generated via spontaneous parametric down-conversion (SPDC) \cite{KMW95,AB00}, a second-order nonlinear optical process that is tantamount to time-reversed sum-frequency (SF) generation \cite{B08_3, VRB20}, and which conserves both spin and OAM. However, the generation and manipulation of entangled light is hindered not only by the low nonlinear response of conventional materials, but also by the need to collect and direct the entangled photon pairs---produced upon phase-matching in separate bulk nonlinear crystals---into scalable optical components that enable quantum logic operations. Theoretical explorations of SPDC by waveguided photons have revealed its feasibility in the presence of material dispersion and loss \cite{YLS08,HSS15}, while experimental efforts to develop on-chip sources of entangled photons include demonstrations of SPDC in periodically poled LiNbO$_3$ waveguides \cite{PSB09,LHK15} and in a microring resonator \cite{ISM04,GZS17}. Additionally, the SPDC process has been recently proposed to conserve the in-plane momentum in graphene ribbons containing an electrostatically induced p-n junction where plasmonic modes are entangled \cite{SBF21}.

% Figure 1 ------------------------------------------------
\begin{figure*}
\centering \includegraphics[width=1.0\textwidth]{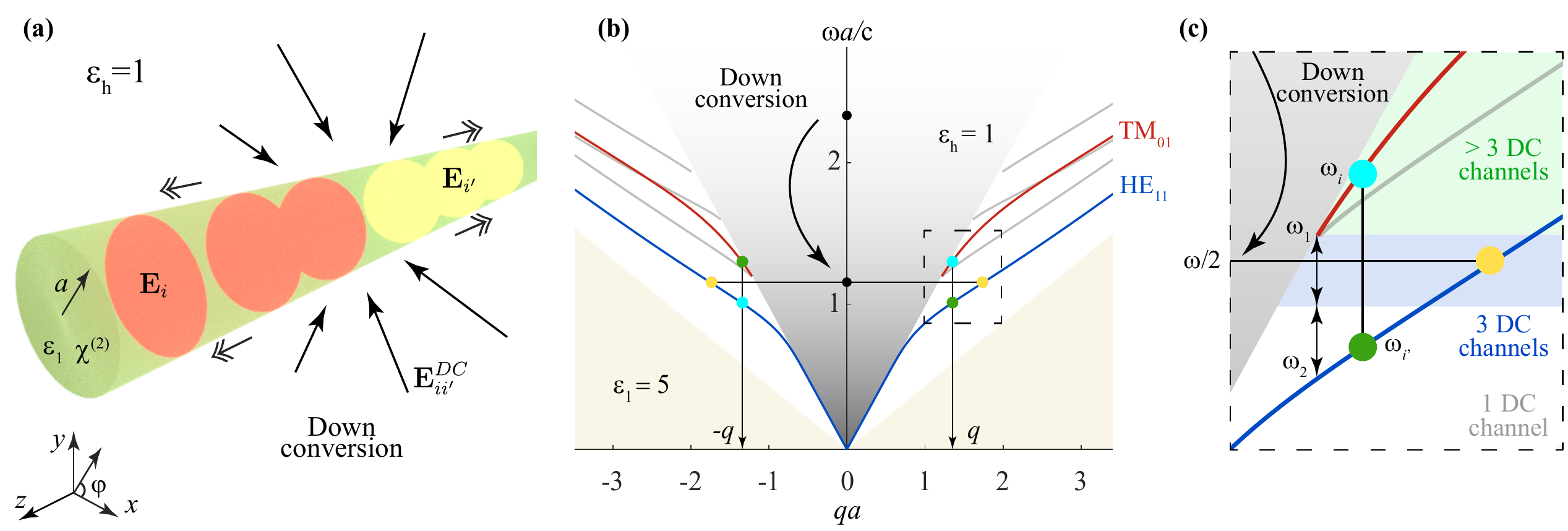}
\caption{{\bf Generation of waveguided entangled photon pairs by down-conversion in an optical fiber.} {\bf (a)} Illustration of a cylindrical fiber (radius $a$, material permittivity $\epsc$, host permittivity $\epsh$) subject to normal illumination. Each incident photon can be down-converted via the second-order nonlinear response of the fiber material (susceptibility $\chi^{(2)}$) to produce two waveguided photons within modes $i$ and $i'$ of fields $\Eb_i$ and $\Eb_{i'}$, frequencies $\omega_i$ and $\omega_{i'}$, and wave vectors $q_i$ and $q_{i'}$ satisfying $q_i+q_{i'}=0$. {\bf (b)} Dispersion diagram of waveguide modes (normalized frequency $\omega a/c$ as a function of normalized wave vector $qa$) for $\epsc=5$ and $\epsh=1$. The light cones in the waveguide and host materials (white and grey areas, respectively) limit the existence of the modes. We highlight the two lowest-order modes that possess nonzero longitudinal field components (HE$_{11}$ and TM$_{01}$, see labels) and enable down-conversion with a small number of emission photon-pair channels: one symmetric (yellow circles) and two asymmetric (blue and green circles) channels. {\bf (c)} Detail of these channels, showing the threshold frequency of the TM$_{01}$ mode $\omega_1$, the frequency $\omega_2$ of the HE$_{11}$ mode with the same wave vector, and the number of down-conversion channels available depending on the incident photon frequency $\omega$ (1, 3, and $>3$ in white blue and green areas). Each channel has $\pm q$ and $\pm m$ degeneracies.}
\label{Fig1} 
\end{figure*}

% Figure 2 ------------------------------------------------
\begin{figure*}
\centering \includegraphics[width=0.5\textwidth]{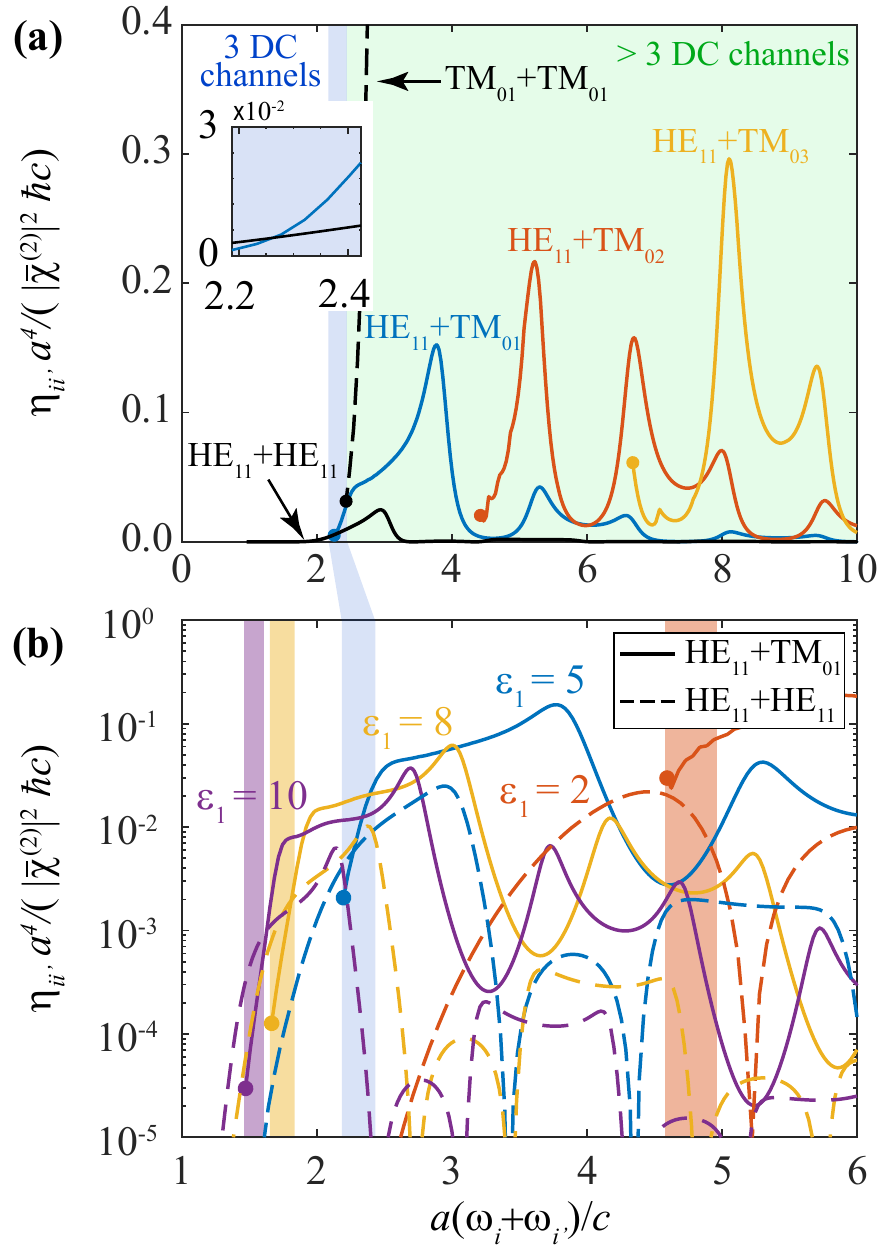}
\caption{{\bf Down-conversion efficiency for different output channels.} {\bf (a)} Normalized SPDC efficiency for a waveguide with $\epsc=5$ and $\epsh=1$ as a function of incident light frequency $\omega=\omega_i+\omega_{i'}$. Different output channels $i+i'$ are indicated by labels. Areas highlighted in white, blue, and green correspond to 1, 3, and $>3$ available channels (each of them degenerate in the sign of both the wave vectors and the azimuthal numbers). {\bf (b)} Efficiencies corresponding to the three lowest channels (one in HE$_{11}$+HE$_{11}$ and two in HE$_{11}$+TM$_{01}$) for different values of the waveguide permittivity $\epsc$ (see color-coded labels). Colored areas highlight the respective regions in which three channels exist, while the frequency threshold for each of the down-conversion channels is indicated by colored circles.}
\label{Fig2} 
\end{figure*}

% Figure 3 ------------------------------------------------
\begin{figure*}
\centering \includegraphics[width=0.4\textwidth]{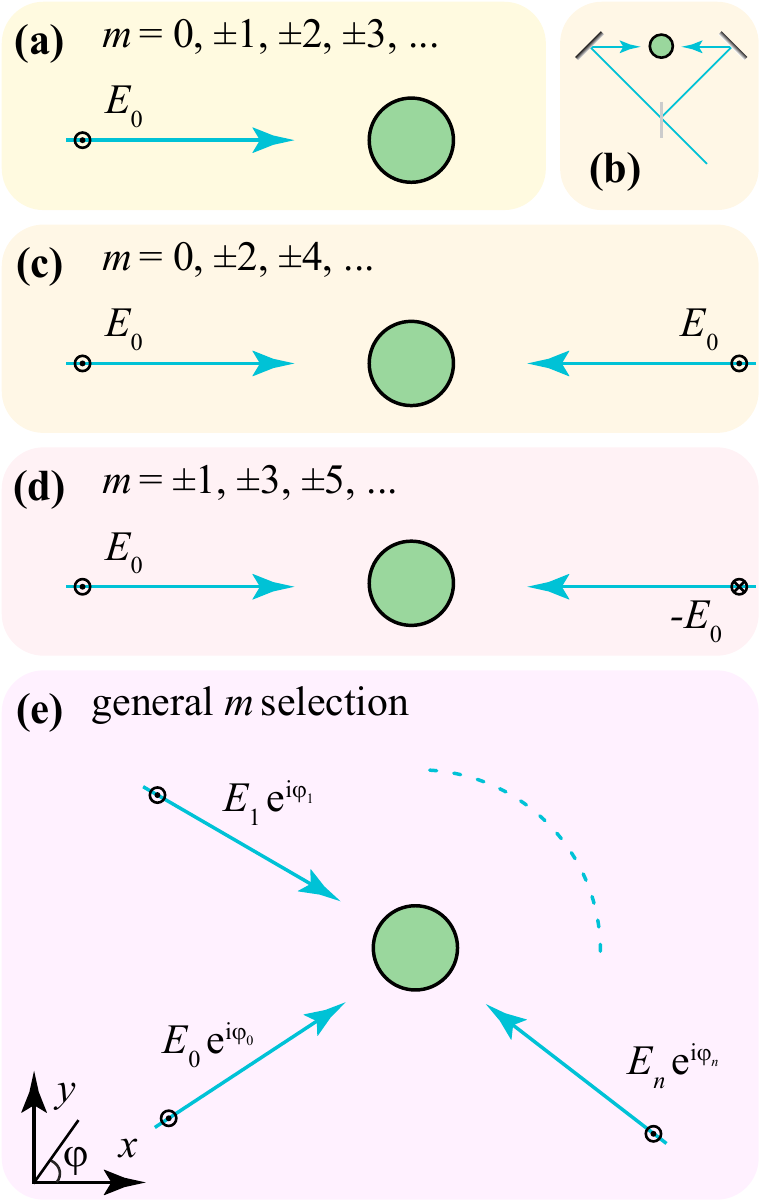}
\caption{{\bf Mode selection through light interference.} A single incident light plane wave (a) contains all possible values of the azimuthal number $m$ in the external field, and can thus excite all available SPDC channels for the chosen input frequency. We can combine illumination from different directions through beam splitters and mirrors, as indicated in (b) for two-plane-wave irradiation, leading to a selection of the $m$ values that remain in the external light. Examples of selection by irradiation with two in-phase and out-of-phase counter-propagating plane waves are shown in (c) and (d). More stringent selection of $m$ is possible by combining multiple plane waves of amplitudes $E_j$ along different azimuthal directions $\varphi_j$, with $j=0,\dots,n$.}
\label{Fig3} 
\end{figure*}

In this work, we propose an alternative strategy to excite entangled photon pairs directly into a low-loss optical waveguide simply by illuminating the waveguide from free space, and explore the feasibility of this approach through rigorous theoretical analysis. Our method relies on the intrinsic second-order optical nonlinearity of the waveguide to down-convert a normally impinging optical field directly into two guided modes, where energy and momentum conservation restricts the possible modes that can be accessed by a particular incident field. To quantitatively analyze the down-conversion scheme, we consider the reverse process, in which two counter-propagating waveguide modes up-convert into a free-space photon mode. By invoking the reciprocity theorem \cite{NH06}, our analysis effectively describes the fidelity of our proposed SPDC scheme, which can be readily explored in an experimental setting using conventional optical components, and thus provides a widely accessible source of entangled photon pairs directly generated in an optical fiber. Although different counter-propagating illumination schemes have been proposed \cite{DB02,BAD02,OCL11,SPS17}, we emphasize here that entanglement takes place directly within the modes of the optical fiber.

% =========================================================
% --- results ---------------------------------------------
% =========================================================
\section{Results and discussion}  \label{sec:Results}

We consider the configuration shown in Figure\ \ref{Fig1}(a), consisting of a freestanding cylindrical waveguide of radius $a$ under normal illumination. For simplicity, we assume isotropic, homogeneous materials, although our calculations can be readily extended to anisotropic media and more complex geometries, such as a noncircular waveguide on a substrate. The second-order nonlinearity of the waveguide material facilitates SPDC into states lying within different bands, so that an incident photon is converted into two photons guided along the waveguide, with wave vectors of opposite sign ($q$ and $-q$) in order to conserve momentum along the direction of translational invariance, as sketched in Figure\ \ref{Fig1}(b) and discussed below.

Without entering into the details of how to quantify entanglement for more complex states \cite{P14}, we aim at producing maximally entangled photon pairs moving along opposite directions along the waveguide (left $L$ with wave vector $-q$, and right $R$ with wave vector $q$) that correspond to Bell quantum states of the form
\begin{align}
|\psi\rangle=|L_iR_{i'}\rangle+|L_{i'}R_i\rangle \label{psiLR},
\end{align}
where $i$ and $i'$ denote different photon quantum numbers, such as the azimuthal number $m$, the mode polarization, and the frequency. Before exploring these possibilities, we provide a rigorous theory to calculate the SPDC efficiency associated with different output channels in the waveguide.

\subsection{Down-conversion efficiency in cylindrical waveguides}
\label{theory}

To quantify the SPDC efficiency, we compute the probability of the inverse process: SF generation produced by two counter-propagating guided photons of frequencies $\omega_i$ and $\omega_{i'}$, which are combined to generate a photon of frequency $\omega_{ii'}=\omega_i+\omega_{i'}$ that is normally emitted from the waveguide. In virtue of reciprocity, the per-photon probabilities for the two processes (SPDC and SF generation) are identical. In practice, we calculate the efficiency by considering two photons within counter-propagating guided modes $i$ and $i'$, prepared as long pulses of length $L$ and space/time-dependent electric fields $\Eb_i(\rb,t)$ and $\Eb_{i'}(\rb,t)$ (Figure\ \ref{Fig1}(a)) that comprise frequency components that are tightly packed around $\omega_i$ and $\omega_{i'}$. Through the SF second-order susceptibility tensor $\chi^{(2)}$, a polarization density $\Pb_{ii'}(\Rb)$ is produced within a narrow frequency range around $\omega_{ii'}$. More precisely,
\begin{align} \label{Eq:P_a_auxbis}
    \tilde{P}_{ii',a}(\Rb) = \sum_{bc} \frac{\chi^{(2)}_{abc}(\Rb)}{|\bar\chi^{(2)}|} E_{i,b}(\Rb)   E_{i',c}(\Rb),
\end{align}
where the indices $\{a,b,c\}$ run over Cartesian components, $\Eb_i(\Rb)$ gives the profile of mode $i$ in the transverse plane $\Rb=(x,y)$, and we normalize the susceptibility to the quantity
\begin{align} \label{Eq:chi2_average}
    |\bar\chi^{(2)}|\equiv\sum_{abc}|\chi^{(2)}_{abc}|.
\end{align}
For simplicity, we consider the wave vectors of the two modes to satisfy the condition $q_i+q_{i'}=0$, so that the SF photons are emitted with zero wave vector along the waveguide (i.e., along normal directions). The SF polarization density generates a field that we compute at long distances from the fiber using the electromagnetic Green tensor of the system $\Gg(\rb,\rb',\omega)$, from which we calculate the far-field Poyniting vector, whose radial component is in turn integrated over time and directions of emission to produce the emitted energy. We then divide this energy by $\hbar\omega_{ii'}$ to obtain the number of emitted photons $N_{ii'}$. Likewise, we calculate the Poynting vector associated with each of the pulses and integrate the component parallel to the waveguide over time and transverse spatial directions to yield the number of photons incident in each pulse, $N_i$ and $N_{i'}$. Finally, the ratio of the emitted number of photons to the number of photons in each pulse is interpreted as the probability $\eta_{ii'}=N_{ii'}/N_iN_{i'}$ that two colliding quanta combine into one emitted SF quantum (again, identical with the probability that an externally incident photon produces a pair of counter-propagating photons within modes $i$ and $i'$). In the long $L$ limit, the incident pulses become monochromatic and $\eta_{ii'}$ turns out to be independent of $L$. For convenience, we separate the up-conversion efficienty into the contributions associated with the emission along different azimuthal angles $\varphi$ (see coordinate system in Figure\ \ref{Fig1}(a)) as
\begin{align}
\eta_{ii'}=\int_0^{2\pi}\eta_{ii'}(\varphi)\,d\varphi. \label{eta}
\end{align}
After a lengthy calculation (see a detailed self-contained derivation in Methods), we find the following result for the angle-resolved efficiency:
\begin{widetext}
\begin{subequations} \label{etafinal}
\begin{align}
\eta_{ii'}(\varphi)&=\frac{2\pi\hbar c}{a^4} |\bar\chi^{(2)}|^2 \frac{w_iw_{i'}}{(w_i+w_{i'})^2} \frac{|\beta_i\beta_{i'}|}{\vert \beta_{i}\vert+\vert\beta_{i'} \vert} \frac{I_{ii'(\varphi)}}{I_iI_{i'}},
\label{etaphi}\\
I_i &=\frac{1}{a^2}\int d^2\Rb \;  {\rm Re}\left\{E_{i,x}(\Rb) H_{i,y}^*(\Rb)- E_{i,y}(\Rb) H^*_{i,x}(\Rb)\right\},
\label{Eq:I_i} \\
I_{ii'}(\varphi) &=\left|\;\int_{R'<a}\!\!\! d^2\Rb' \, \gb\left(\varphi-\varphi',R',\omega_{ii'}\right) \cdot\tilde\Pb_{ii'}(\Rb')\right|^2,
\label{Eq:I_ii}
\end{align}
\end{subequations}
\end{widetext}
where $w_i=\omega_{i}a/c$, $\beta_i=v_i/c$, $v_i=\partial\omega_i/\partial q_i$ is the group velocity in mode $i$, and $\gb(\varphi-\varphi',R',\omega)$ is the far-field-limit amplitude of the electromagnetic Green tensor defined though $\Gg(\rb,\rb',\omega) \rightarrow (\ee^{\ii\sqrt{\epsh}\omega R/c}/R)\,\gb(\varphi-\varphi',R',\omega)$ for normal emission (see Eq.\ \eqref{gexplicit} below for an explicit expression). Here, $I_i$ and $I_{ii'}$ are proportional to the number of photons incident within the waveguide mode $i$ and emitted outside the waveguide, respectively. Incidentally, these coefficients are normalized in such a way that they are independent of the waveguide radius $a$, so the efficiency $\eta_{ii'}$ only depends on $a$ through an overall factor $1/a^4$ for a fix value of $\omega_ia/c$. In brief, $\eta_{ii'}$ represents the ratio of SF photons produced per two incident photons (one in each waveguide mode), that is, the SF matrix element for $\omega_{i}+\omega_{i'} \to \omega_{ii'}$, which must be equal to the SPDC matrix element corresponding to $\omega_{ii'} \to \omega_{i}+\omega_{i'}$. The latter affects each incident photon separately, so it can be interpreted as the fraction of incident photons that undergo SPDC, and therefore, also the fraction of down-converted power.

For the cylindrical waveguides under consideration, we can multiplex the mode labels as $i=\{q_i,m_i,l_i,\sigma_i\}$, where $q_i$ is the wave vector, $m_i$ is the azimuthal angular momentum number, $l_i$ refers to different radial resonances, and $\sigma_i$ runs over polarization states (i.e., TE$_{0l_i}$ and TM$_{0l_i}$ for $m_i=0$, and hybrid modes HE$_{m_il_i}$ and EH$_{m_il_i}$ for $m_i\neq0$, see Sec.\ \ref{Sec4.1.1}). Given the symmetry of the waveguide, the radial and azimuthal components of the transverse field associated with each mode only depend on radial distance $R$, apart from an overall phase factor $\ee^{\ii m_i\varphi}$. For simplicity, we consider a second-order response tensor $\chi^{(2)}$ that also preserves the cylindrical symmetry, so that the angular integral in Eq.\ \eqref{Eq:I_ii} leads to angular momentum conservation ($m_{ii'}=m_i+m_{i'}$ for the emitted photons). In particular, we assume a nonlinear tensor dominated by the $\chi^{(2)}_{zzz}$ component (e.g., a LiNbO$_3$ waveguide with the $z$ axis aligned along the waveguide). This implies that the TE modes and the TE component of the hybrid modes do not couple to the incident field through $\chi^{(2)}$.

\subsection{Availability and efficiency of different down-conversion channels}
\label{avail}

We are now equipped to discuss the generation of entangled photon pairs through SPDC in our waveguide. Assuming the above conditions, the lowest-frequency modes that possess a nonzero $z$ field, and can consequently couple to normally impinging external light, are HE$_{11}$ and TM$_{01}$ (see Figure\ \ref{Fig1}(b)). We identify two relevant frequencies in this region (see Figure\ \ref{Fig1}(c)): the threshold of the TM$_{01}$ mode at $\omega_1$ (satisfying $\omega_1a/c=2.4048/\sqrt{\epsh-\eps_1}$, see Sec.\ \ref{Sec4.1}); and the frequency $\omega_2$ of mode HE$_{11}$ with the same wave vector. Upon inspection, we find that for an incident light frequency $\omega<\omega_1+\omega_2$, the only SPDC channel that is available corresponds to the generation of two HE$_{11}$ modes of frequency $\omega/2$ and opposite wave vectors. This situation already allows us to produce entangled photon pairs of the form given in Eq.\ \eqref{psiLR}, where $i$ and $i'$ now refer to the azimuthal numbers $m_i,m_{i'}\in\{-1,1\}$ in each of the emitted photons. In particular, if the fiber is symmetrically illuminated along different azimuthal directions (see Sec.\ \ref{selection} below), it is possible to select only the $m=0$ component from the external light, so that conservation of azimuthal angular momentum leads to the condition $m_i+m_{i'}=0$, and therefore, the emitted photon pair forms an entangle state $|L_{-1}R_{1}\rangle+|L_{1}R_{-1}\rangle$, where the subindices indicate the values of $m_i$ and $m_{i'}$ for the $L$ and $R$ emission directions, all of them sharing the same frequency $\omega/2$ and polarizations HE$_{\pm1,1}$, so we refer to these channel as HE$_{11}$+HE$_{11}$.

Another interesting range of incidence frequencies is $\omega_1+\omega_2<\omega<2\omega_1$ (blue area in Figure\ \ref{Fig1}(c)), where the HE$_{11}$+HE$_{11}$ channel is now supplemented by two additional possibilities in which the two generated photons have different frequencies (with the sum satisfying $\omega=\omega_i+\omega_{i'}$) and lie in different bands (HE$_{11}$ or TM$_{01}$). This is indicated by the two pairs of color-matched blue and green dots in Figure\ \ref{Fig1}(b)-(c), where the condition of opposite wave vectors is obviously satisfied. Again, it is possible to select a specific SPDC channel by illuminating with a fixed $m$ number (see below), and in particular, by setting $m=m_i+m_{i'}=1$, the HE$_{11}$+HE$_{11}$ channel is eliminated (because the overall azimuthal number obtained by combining two HE$_{\pm11}$ modes is 0 or $\pm2$), so that we obtain again a maximally entangled state of the form given in Eq.\ \eqref{psiLR} with $i$ and $i'$ now referring to TM$_{01}$ and HE$_{11}$ (i.e., $|L_{\rm TM}R_{\rm HE}\rangle+|L_{\rm HE}R_{\rm TM}\rangle$, with azimuthal numbers $m_i$ and $m_{i'}$ taking the values 0 and 1 in the TM and HE components, respectively).

The formalism presented in Sec.\ \ref{theory} allows us to calculate the SPDC efficiency for the production of specific photon pair states, with external illumination prepared with an azimuthal number $m=m_i+m_{i'}$ and a polarization state determined by the time reversal of the sum-frequency generation state considered in the derivation of these results. Under the assumed conditions of incidence along transverse directions, and considering a $zzz$ dominant component in the second-order susceptibility tensor, the profile of the applied light amplitude as a function of azimuthal angle $\varphi$ is therefore taken to be $\ee^{\ii m\varphi}$, with the field oriented parallel to the waveguide direction. These conditions can be met by combining several incident light beams, as we discuss below. The efficiencies calculated for different SPDC channels in this scheme are shown in Figure\ \ref{Fig2}, normalized to $(\hbar c/a^4)|\bar\chi^{(2)}|^2$ in order to present universal, dimensionless results as a function of the scaled incident light frequency $\omega a/c$. For an $\epsc=5$ waveguide in air (Figure\ \ref{Fig2}(a)), we find efficiencies that generally grow with the order of the waveguide modes, exhibiting resonances as a function of the incident frequency $\omega=\omega_i+\omega_{i'}$. These resonances are inherited from the two-dimensional transmission coefficients (see Sec.\ \ref{Sec4.1}) and can be understood as coupling of the incident light to leaky cavity modes at the incident light frequency. We indicate in white the area in which there is only one decay channel (HE$_{11}$+HE$_{11}$, see above), whereas the area with three decay channels (two additional ones corresponding to TM$_{01}$+HE$_{11}$ and HE$_{11}$+TM$_{01}$) is highlighted in blue. The green region at higher frequencies contains an increasing number of channels, which could be also exploited to generate more complex entangled mixtures, involving multiple output states in each direction ($L$ and $R$) and higher-order modes.

The spectral evolution of the efficiencies is roughly maintained when varying the waveguide permittivity $\epsc$ (Figure\ \ref{Fig2}(b)), but we observe a general increase in $\eta_{ii'}$ with increasing $\epsc$ in the region of interest, as well as a spectral shift of the region with three output channels (highlighted in shading colors and evolving toward lower frequencies as we increase the permittivity, in agreement with the single-mode-fiber cutoff condition). Interestingly, we find a crossover is the efficiency of HE$_{11}$+HE$_{11}$ relative to that of HE$_{11}$+TM$_{01}$: the first one dominates over the second one within the three-channel region at high $\epsc$, whereas the opposite behavior is found at lower permittivities.

These numbers indicate that the current scheme is feasible for producing a reasonable rate of entangled photon pairs, taking into account that they are already prepared within waveguide modes \cite{DB02,OCL11}. In particular, for values of $|\bar\chi^{(2)}|\sim10^{-10}\,$m/V found in good nonlinear materials such as LiNbO$_3$ \cite{DGN99,B08_3} and a waveguide radius $\sim100$\,nm, the scaling factor in Figure\ \ref{Fig2} is $(\hbar c/a^4)|\bar\chi^{(2)}|^2\sim10^{-10}$, which yields a power fraction of $10^{-11}$ for SPDC when it is multiplied by a scaled efficiency of $\sim0.1$ (Figure\ \ref{Fig2}). Considering photon energies $\sim1\,$eV and an incident light power $\sim1\,$mW, this amounts to a generation rate of $\sim10^5$ entangled photon pairs per second. As an additional possibility, the efficiency could be increased by incorporating resonant elements to amplify the external light in the region surrounding the waveguide, such as planar Fabry-Perot resonators, which is a natural option for waveguides fabricated on a substrate.

\subsection{Selection of down-conversion channels through illumination interference}
\label{selection}

A $p$-polarized light plane wave of amplitude $E_0$ incident through the host medium with a wave vector $\kb_{\rm h}\perp\zz$ normal to the waveguide (Figure\ \ref{Fig3}(a)) contributes with a broad range of azimuthal numbers $m=m_i+m_{i'}$ according to the decomposition
\begin{align}
E_0\,\zz\,\ee^{\ii\kb_{\rm h}\cdot\Rb}=\sum_m \ii^{-m}\, E_0 \ee^{-\ii m \varphi_{\kb_{\rm h}}} \Eb^J_{{\rm h},0,mp}(\Rb) \label{E0field}
\end{align}
in terms of cylindrical waves $\Eb^J_{{\rm h},0mp}$ (see Sec.\ \ref{planetocyl}). This situation leads to entangled states that combine more than two polarizations for each of the two waveguiding directions ($L$ and $R$). For example in the single HE$_{11}$+HE$_{11}$ channel region (at incident light frequency $\omega<\omega_1+\omega_2$), we can have all combinations of $m_i=\pm1$ and $m_{i'}=\pm1$, thus reducing the degree of entanglement. A way to fix this problem is by combining illumination from different azimuthal directions (e.g., in an interferometric setup involving beamsplitters and mirrors, as illustrated in Figure\ \ref{Fig3}(b)). In particular, when illuminating with two in-phase counter-propagating waves (Figure\ \ref{Fig3}(c)), only even values of $m$ survive, whereas only odd $m$'s are selected if the waves have a $\pi$ relative phase difference (Figure\ \ref{Fig3}(d)). In general, we can consider an arbitrary number of plane waves (Figure\ \ref{Fig3}(e)), so that, the total field acting on the waveguide has the same form as in Eq.\ \eqref{E0field}, but with $E_0\ee^{-\ii m \varphi_{\kb_{\rm h}}}$ substituted by $\sum_jE_j\ee^{-\ii m \varphi_j}$, where the sum runs over plane waves $j$ of amplitude $E_j$, directed along azimuthal directions $\varphi_j$.

In the one-channel region (HE$_{11}$+HE$_{11}$ output), we can generate the maximally entangled state $|L_{-1}R_{1}\rangle+|L_{1}R_{-1}\rangle$ by just selecting an incident $m=0$ and eliminating $m=\pm2$, as other values of $m$ do not couple to the output modes in that region (e.g., with equal amplitudes and azimuthal angles of $0$ and $\pm \pi/3$). This selection requires a minimum of three external plane waves. Likewise, in the three-channel region, only three plane waves are required to select $m=1$ (or $m=-1$) while discarding the undesired $m=0,\pm2$ and $m=-1$ (or $m=1$) possibilities and obtain a maximally entangled state $|L_{\rm TM}R_{\rm HE}\rangle+|L_{\rm HE}R_{\rm TM}\rangle$ with $TM_{01}$ and $HE_{11}$ (or $HE_{-11}$) components. A more stringent selection of $m$ is possible by resorting to more incident plane waves, therefore opening a vast range of possible entangled photon pairs prepared in higher-order modes.

% =========================================================
% --- conclusions -----------------------------------------
% =========================================================
\section{Concluding remarks}

We propose a straightforward approach to generate entangled photon pairs directly into low-loss dielectric waveguides based on down-conversion of normally impinging light and introduce a theoretical formalism based on the reciprocity theorem to quantify the efficiency of the process. This formalism leads to a universal overall scaling of the efficiency $\eta$ with the second-order nonlinear susceptibility $\chi^{(2)}$ and waveguide radius $a$ as $\eta\propto\vert\bar{\chi}^{(2)}\vert^2/a^4$, which is further factored by an involved interplay among material parameters. For a moderate incident light power of 1\,mW and an efficient nonlinear material such as LiNbO$_3$, we predict a production rate of $\sim10^{5}$ entangled photon pairs per second. The theoretical prescription here presented for cylindrical geometries can be readily extended to other waveguide configurations, which impose different symmetries. In particular, preferential elements of the nonlinear susceptibility tensor may be more easily accessed in alternative morphologies depending on the material symmetry. Integration of the waveguide on a substrate opens additional possibilities to resonantly amplify the external light (e.g., through Fabry-Perot resonators), including the exposure to evanescent fields (along the transverse directions), rather than propagating light. Crucially, the investigated strategy to generate counter-propagating photons necessitates only conventional optical elements, while the theory can be directly applied to predict the efficiency of the down-conversion process. Moreover, we suggest interferometric schemes to select the symmetry of the generated photon modes, thus reducing the number of accessible SPDC channels and increasing the resulting degree of entanglement. Frequency post-selection of the generated waveguided photons can also be used to discard undesired channels and enhance entanglement. An implementation of these ideas should enable the generation of down-converted photon pairs with a high degree of entanglement involving on-demand combinations of high-order symmetries. We thus envision that these findings can stimulate experimental ventures in quantum optics to entangle light with a predictable degree of fidelity and help alleviate practical issues related to the coupling of quantum light sources to optical components required in emerging quantum information technologies.

% =========================================================
% --- methods/appendix ------------------------------------
% =========================================================
\section{Methods} \label{sec:Methods} %---ACS optional
\begin{widetext} %---arxiv
%\appendix %---APS---OSA---SI---arxiv optional
%\renewcommand{\thesection}{A} %---SI---arxiv optional
%\renewcommand{\theequation}{A\arabic{equation}} %---SI---arxiv optional
%\setcounter{equation}{0} %---OSA optional

% --- text for methods/appendix ----------------------------
%\noindent{\bf Title Format.} ... text ... %---ACS format for Methods section headers
%\section{title} ... text ... %---APS---SI---arxiv format for Appendix section headers (not for OSA)
%\subsection{title} ... text ... %---APS---OSA---SI---arxiv format for Appendix subsection headers
%\renewcommand{\theequation}{A\arabic{equation}} %---OSA for A, B, ... subsections of Appendix

In this section, we provide a detailed, self-contained derivation of the formalism and equations used in the main text. More precisely, we provide the following elements: a description of guided modes in a cylindrical dielectric wire, along with explicit expressions for their associated electromagnetic fields; a discussion of waveguided pulses; a study of the field produced by line dipoles situated inside the waveguide; a calculation of the SF energy that is emitted into the far field through the second-order nonlinearity of the waveguide material in response to two counter-propagating guided pulses; and a derivation of the SF conversion efficiency, which we argue to be equal to the SPDC efficiency in virtue of reciprocity. We consider guided modes with opposite wave vectors (Figure\ \ref{Fig1}(a)), which couple to external light propagating along directions perpendicular to the waveguide.

% =========================================================
\subsection{Electromagnetic waves in a cylindrical waveguide}
\label{Sec4.1}

To describe electromagnetic waves in a cylindrical geometry, we first decompose the electric field into cylindrical waves following the prescription of Ref.\ \cite{paper047}. More specifically, adopting a cylindrical coordinate system $\rb=(R,\varphi,z)$, we consider a homogeneous, isotropic dielectric medium (labeled $j$) free of external charges and currents that is characterized by a permittivity $\eps_j$ (setting the magnetic permeability to $\mu=1$) and express the electric field in cylindrical waves indexed by their azimuthal number $m$, wave vector $q$ along $\zz$, and polarization $\sigma \in \{s,p\}$ according to
\begin{subequations} \label{Eq:Es_Ep}
\begin{align}
    \Eb^J_{j,qms} (\rb) =& \left[ \frac{\ii m }{Q_j R} J_m(Q_j R) \RR-J_m'(Q_jR)\hat{\varphi} \right] \ee^{\ii m \varphi} \ee^{\ii q z}, && \text{$s$ waves},  \label{Eq:Es} \\
    \Eb^J_{j,qmp}(\rb) =& \frac{q}{k_j} \left[ \ii J_m'(Q_j R) \RR-\frac{m}{Q_j R} J_m(Q_j R) \hat{\varphi}+\frac{Q_j}{q} J_m(Q_j R) \zz \right] \ee^{\ii m \varphi} \ee^{\ii q z}, && \text{$p$ waves,}  \label{Eq:Ep} 
\end{align}
\end{subequations}
where we define $ k_j = \sqrt{\eps_j}\omega/c$ and $ Q_j = \sqrt{k_j^2-q^2+\ii 0^+} $ (with the square root yielding a positive real part), while the primes on the Bessel functions denote differentiation with respect to the argument. From the orthogonality of the Bessel functions ($\int_0^\infty xdxJ_m(x)J_m(ax)=\delta(a-1)$), it is easy to show that these fields satisfy the orthonormality relation $\int d^2\Rb \, \Eb^J_{j,qm\sigma} \cdot \left(\Eb^J_{j,q'm'\sigma'} \right)^* = 2\pi \delta_{mm'} \delta_{\sigma \sigma'}\delta(q-q')/q$, while the field of modes with different polarizations are related as
\begin{align} \label{Eq:s_p_transformation}
    \Eb^J_{j,qm\sigma} = \frac{1}{k_j} \nabla \times \Eb^J_{j,qm\sigma'}, \quad  \quad  \sigma \neq \sigma'.
\end{align}
We now discuss a cylindrical wave emanating from the interior of a cylindrical waveguide of radius $a$ that is infinitely extended in the $z$ direction, comprised of a dielectric material of permittivity $\eps_1$ (medium $j=1$), and embedded in a host medium $j=$h (permittivity $\eps_{\rm h}$). Using the notation introduced above, the electric field is expressed as
\begin{align} \label{Eq:Scattering}
    \Eb = \begin{cases} \Eb^H_{1,q m \sigma}+r_{m,s\sigma} \Eb^J_{1,qms} +  r_{m,p\sigma} \Eb^J_{1,qmp},& \ R<a, \\ 
    t_{m,s \sigma}\Eb^H_{{\rm h},qms} + t_{m,p\sigma} \Eb^H_{{\rm h},qmp},& \ R\geq a,
    \end{cases}
\end{align}
where
\begin{subequations} \label{EqH:Es_Ep}
\begin{align}
    \Eb^H_{j,qms} (\rb) =& \left[ \frac{\ii m }{Q_j R} H^{(1)}_m(Q_j R) \RR-H^{(1)'}_m(Q_jR)\hat{\varphi} \right] \ee^{\ii m \varphi} \ee^{\ii q z}, && \text{$s$ waves},  \label{EqH:Es} \\
    \Eb^H_{j,qmp}(\rb) =& \frac{q}{k_j} \left[ \ii H^{(1)'}_m(Q_j R) \RR-\frac{m}{Q_j R} H^{(1)}_m(Q_j R) \hat{\varphi}+\frac{Q_j}{q} H^{(1)}_m(Q_j R) \zz \right] \ee^{\ii m \varphi} \ee^{\ii q z}, && \text{$p$ waves,}  \label{EqH:Ep} 
\end{align}
\end{subequations}
are outgoing waves similar to the propagating waves in Eqs.\ \eqref{Eq:Es_Ep}, but with the Bessel functions $J_m$ substituted by Hankel functions $H_m^{(1)}$, while the reflection and transmission coefficients $r_{m,\sigma\sigma'}$ and $t_{m,\sigma\sigma'}$ are given by (see Sec.\ \ref{tpts})
\def\1{\\[8pt]}
\begin{equation}  \label{Eq:RT_coefficients}
    \left[ 
        \begin{matrix}
            r_{m,ss} \1 t_{m,ss} \1 r_{m,ps} \1 t_{m,ps}
        \end{matrix}
    \right]
    = M^{-1}
        \left[ 
        \begin{matrix}
           -\zeta \dfrac{ Q_1}{k_1}H^{(1)}_{m}(\tilde{Q}_1) \1 
            -H^{(1)'}_{m}(\tilde{Q}_1) \1
           0 \1 
           -\zeta \dfrac{m q}{k_1 \tilde{Q}_1} H^{(1)}_{m} (\tilde{Q}_1)
        \end{matrix}
    \right]
\quad \quad 
\text{and}
\quad \quad 
    \left[ 
        \begin{matrix}
            r_{m,sp} \1 t_{m,sp} \1 r_{m,pp} \1 t_{m,pp}
        \end{matrix}
    \right]
    = M^{-1}
        \left[ 
        \begin{matrix}
          0  \1
          -\dfrac{m q}{k_1 \tilde{Q}_1} H^{(1)}_{m} (\tilde{Q}_1) \1 
          -\dfrac{ Q_1}{k_1}H^{(1)}_{m}(Q_1 a) \1 
          -\zeta H^{(1)'}_{m}(\tilde{Q}_1) 
        \end{matrix}
    \right],
\end{equation}
with the matrix $M$ defined as \cite{paper047}
\begin{equation}  \label{Eq:M_matrix}
M = 
    \left[
        \begin{matrix}
            \dfrac{\zeta Q_1}{k_1} J_m(\tilde{Q}_1)    & \dfrac{-\Qh}{\kh} H_m^{(1)} (\tQh)     & 0                             & 0                                      \1
            J_m'(\tilde{Q}_1)                        & -H_m^{(1) '}(\tQh)                     & \dfrac{m q}{k_1 \tilde{Q}_1} J_m(\tilde{Q}_1) & \dfrac{-m q}{\kh \tQh} H_m^{(1)}  (\tQh) \1
            0                                & 0                                     & \dfrac{Q_1}{k_1} J_m(Q_1 a)     & \dfrac{-\Qh}{\kh} H_m^{(1)} (\tQh)      \1
            \dfrac{\zeta m q}{k_1 \tilde{Q}_1}J_m(\tilde{Q}_1) & \dfrac{-m q}{\kh \tQh} H_m^{(1)} (\tQh) & \zeta J_m'(\tilde{Q}_1)                 & -H_m^{(1) '} (\tQh) 
         \end{matrix}
    \right],
\end{equation} 
$\tilde{Q}_j=Q_ja$, and $\zeta \equiv \sqrt{\eps_1 / \eps_{\rm h}}$. These expressions are obtained by imposing the electromagnetic boundary conditions at $R=a$, specifically the continuity of the $\hat{\varphi}$ and $\zz$ electric and magnetic field components at the cylinder surface, which automatically guarantees the continuity of the electric displacement and the magnetic field along the $\RR$ direction. Incidentally, the dispersion relation for cylindrical waveguide modes is obtained from the condition ${\rm det}\{M\}=0$, which signals the existence of a nontrivial solution in the absence of an external field, and leads to the expression
\begin{align} \label{Eq:mode_disperion}
    \left[\dfrac{1}{\tilde{Q}_1} \frac{J_m'(\tilde{Q}_1)}{J_m(\tilde{Q}_1)} - \dfrac{1}{\tQh}  \frac{H_m^{(1)'}(\tQh)}{H_m^{(1)}(\tQh)} \right] 
    \left[ \dfrac{\epsc}{\tilde{Q}_1} \frac{J_m'(\tilde{Q}_1)}{J_m(\tilde{Q}_1)}-\dfrac{\epsh}{\tQh}  \frac{H_m^{(1) '} (\tQh)}{H_m^{(1)} (\tQh)}\right] 
    =  \left[ \frac{m q }{k}
    \frac{(\tilde{Q}_1)^2-(\tQh)^2}{(\tilde{Q}_1\tQh)^2} \right]^2
\end{align}
with $k=\omega/c$. The above result is equivalent to other {\it textbook} forms of the dispersion relation for cylindrical waveguide modes \cite{A00,K11,M13_1,B15}, typically expressed in terms of modified Bessel functions {\it in lieu} of Hankel functions. The range of wavelengths $\lambda$ for which only a single mode exists is determined by the condition \cite{ST19} $(a/\lambda)\sqrt{\epsc-\epsh}<(\alpha_0/2\pi)=0.3827$, where $\alpha_0$ is the first zero of $J_0$, thus setting a wavelength threshold for the multimode fibers here considered.

% =========================================================
\subsubsection{Electric field distribution of guided modes}
\label{Sec4.1.1}

For convenience, we introduce normalized $s$- and $p$-polarized fields defined as
\begin{subequations} \label{Eq:sp_modes}
\begin{align} \label{Eq:sp_modess}
    \Eb^s_i(\Rb) = \begin{cases} 
        \dfrac{\ii k_1}{\sqrt{\epsc}Q_1} \dfrac{1}{J_m(\tilde{Q}_1)}       \Eb^J_{1,qms}(\Rb) ,\ & R<a, \\
        \dfrac{\ii \kh}{\sqrt{\epsh}\Qh} \dfrac{1}{H^{(1)}_m(\tQh)} \Eb^H_{{\rm h},qms}(\Rb) ,\ & R\ge a,
    \end{cases}
\end{align}
\begin{align} \label{Eq:sp_modesp}
    \Eb^p_i(\Rb) = \begin{cases} 
        \dfrac{k_1}{Q_1} \dfrac{1}{J_m(\tilde{Q}_1)}       \Eb^J_{1,qmp}(\Rb) ,\ & R<a, \\
        \dfrac{\kh}{\Qh} \dfrac{1}{H^{(1)}_m(\tQh)} \Eb^H_{{\rm h},qmp}(\Rb) ,\ & R\ge a,
    \end{cases}
\end{align}
\end{subequations}
respectively, such that $\Eb^{p}_{i}\cdot\zz=1$ and $\Hb^{s}_{i}\cdot\zz=1$, with the magnetic field $\Hb_i^\sigma = -(\ii/k) \nabla\times \Eb_i^\sigma$ obtained from Faraday's law. Note that we evaluate the modes at $z=0$, and a global factor $\ee^{\ii qz}$ is understood to contain the dependence on the coordinate along the waveguide $z$. Guided modes are obtained as solutions of Eq.\ \eqref{Eq:mode_disperion}, which for a given \emph{azimuthal} dependence $m$ admits different \emph{radial} solutions (labeled by $l$), so that the modes are characterized with $\{m,l\}$ indices.

% =========================================================
{\it TE and TM modes.---}For $m=0$ we see from the secular matrix $M$ in Eq.\ \eqref{Eq:M_matrix} that $s$ and $p$ components are not mixed by scattering at the circular waveguide surface, and therefore, pure-polarization solutions exist in this case, signalled by the vanishing of one of the two factors in the left-hand side of Eq.\ \eqref{Eq:mode_disperion}: TE$_{0l}$ modes ($s$ waves) of electric field $\Eb^s_i(\Rb)$ (Eq.\ \eqref{Eq:sp_modess}) when the first factor is zero; and TM$_{0l}$ modes ($p$ waves) of electric field $\Eb^p_i(\Rb)$ (Eq.\ \eqref{Eq:sp_modesp}) when the second factor vanishes.

{\it HE and HE hybrid modes.---}For $m \neq 0$, the solutions to Eq.\ \eqref{Eq:mode_disperion} are modes of hybrid polarization, EH$_{ml}$ and HE$_{ml}$, for which both $s$ and $p$ waves contribute, such that the field of mode $i$ can be expressed as $\Eb_i = \nu \Eb^s_i + \Eb^p_i$, where
\begin{align}
    \nu = \frac{\ii m q}{k} \frac{\tilde{Q}_1^2-\tQh^2}{\tilde{Q}_1\tQh} \left[\tQh J_m'(\tilde{Q}_1)/J_m(\tilde{Q}_1)-\tilde{Q}_1H^{(1)'}_m(\tQh)/H^{(1)}_m(\tQh)\right]^{-1} \nonumber
\end{align}
is defined by imposing continuity of the tangential fields at $R=a$. When $\ii\nu>0$ is far from the cutoff frequency, the modes are termed HE$_{ml}$, while in the opposite situation they are labeled as EH$_{ml}$. Note that alternative equivalent definitions exist depending on how modes are normalized \cite{YS08_1}.

% =========================================================
\subsubsection{Waveguided pulses}

We consider the propagation of Gaussian wavepackets in the cylindrical waveguide, characterized by a finite spatial pulse width $L$ along the waveguide direction $\zz$, such that the field is given by
\begin{align}
    \Eb_i(\rb,t) = \int \frac{dq}{2\pi} \, \Eb_i(\Rb,q) \ee^{\ii (q z-\omega t)} \left[ \sqrt{\pi} L \ee^{-(q-q_i)^2 L^2 / 4} \right] + {\rm c.c.}, \nonumber
\end{align}
where $\Eb_i(\Rb,q)$ is the profile of mode $i$ for a wave vector $q$. In the pulse, $q$ is tightly packed around $q=q_i$. Linearizing the dispersion according to $\omega \approx \omega_i + v_i(q-q_i)$, where $v_i=\partial \omega/\partial q|_{q=q_i}$ is the associated group velocity, and considering $L$ to be large enough to assume that the electric field profile does not vary significantly within a wave vector interval of size $\sim 1/L$ around $q_i$, such that $\Eb_i(\Rb,q)\approx \Eb_i(\Rb,q_i)\equiv\Eb_i(\Rb)$, we can write the field as
\begin{align}
    \Eb_i(\rb,t) \approx \Eb_i(\Rb) \ee^{\ii (q_i z-\omega_i t)} \int \frac{dq}{2\pi} \,  \ee^{\ii (q-q_i)(z-v_it)} \left[ \sqrt{\pi} L \ee^{-(q-q_i)^2 L^2 / 4} \right] + {\rm c.c.}, \nonumber
\end{align}
which, after evaluating the integral in $q$, reduces to
\begin{align} \label{Eq:E_Rt_i}
    \Eb_i(\rb,t) = \Eb_i(\Rb) \ee^{\ii (q_i z-\omega_i t)} \ee^{-(z-v_i t)^2/L^2} + {\rm c.c.}
\end{align}
The corresponding magnetic field is readily computed from Faraday's law $\Hb_i = -(\ii/k) \nabla\times \Eb_i$ by approximating the $\partial_z$ component of $\nabla = \nabla_\Rb+\zz\partial_z$ acting on $\Eb_i$ as $\ii q_i-(z-v_it)/L^2 \approx \ii q_i$, provided that the spatial width of the wavepacket $L$ satisfies $q_i L\gg2\pi $, so that
\begin{align}  \label{Eq:H_Rt_i}
    \Hb_i(\rb,t) = \Hb_i(\Rb) \ee^{\ii (q_i z-\omega_i t)} \ee^{-(z-v_it)^2/L^2} + {\rm c.c.}
\end{align}
with $\Hb_i(\Rb) = -(\ii/k)(\nabla_\Rb+\ii q_i \zz) \times \Eb_i(\Rb)$.

% =========================================================
\subsection{Field produced by an inner line dipole in the region outside the waveguide}
\label{Sec4.2}

We consider a line dipole placed at a transverse position $\Rb_0=(x_0,y_0)$ within the waveguide and represent it by a dipole density $\pb\,\ee^{\ii qz_0}$ (dipole per unit length) extending along the line defined by varying $z_0$ in $(\Rb_0,z_0)$.
%Here, $\pb$ is independent of $z_0$ because we are interested in light emitted normally to the waveguide.
It is useful to begin by calculating the electric field produced in a homogeneous medium with the same permittivity $\epsilon_1$ as the waveguide material, expressed as the integral over $z_0$ of the field due to a point dipole \cite{NH06}:
\begin{align} \label{eq:E_d}
    \Eb^{\rm dip}(\rb,\Rb_0) = \frac{1}{\eps_1}\left(k_1^2 + \nabla \otimes \nabla \right) \pb(\Rb_0) \int dz_0 \, \frac{\ee^{\ii k_1 |\rb-\rb_0|}}{|\rb-\rb_0|}\,\ee^{\ii q z_0},
\end{align}
where $\nabla$ is understood to act on $\rb$, whereas the integral can be evaluated using the identity $\int dz_0\,\exp(\ii k_1 |\rb-\rb_0|+\ii qz)/|\rb-\rb_0| = \ii\pi \ee^{\ii qz} H^{(1)}_0 \left( Q_1|\Rb-\Rb_0|\right)$ with $Q_1$ defined as in Eqs.\ \eqref{Eq:Es_Ep}. The line dipole should generate a set of outgoing cylindrical waves (therefore the Hankel functions) centered at $\Rb_0$, so in order to capitalize the axial symmetry of the waveguide, we need to express the field in terms of waves centered at the origin $\Rb=0$. To this end, we invoke Graf's theorem (see Eq.\ 9.1.79 in Ref.\ \cite{AS1972}), $H_0^{(1)}\left(Q_1 |\Rb-\Rb_0|\right) = \sum_m H_m^{(1)}(Q_1 R) J_m(Q_1 R_0) \,\ee^{\ii m (\varphi-\varphi_0)}$, which holds for $|\Rb| > |\Rb_0|$ and can thus be used to describe the dipole field in the waveguide surface region $R=a>R_0$, through which the outgoing waves are partially transmitted outside the fiber. This translation formula allows us to recast Eq.\ \eqref{eq:E_d} into
\begin{align}\label{Edbis}
    \Eb^{\rm dip}(\rb,\Rb_0) =
    \frac{\ii \pi}{\eps_1} \sum_m J_m(Q_1 R_0) \ee^{-\ii m \varphi_0} \left[k_1^2\pb + \nabla (\pb \cdot \nabla) \right] \, H_m^{(1)}(Q_1 R) \, \ee^{\ii m \varphi}\ee^{\ii qz}.
\end{align}
Now, projecting the dipole as $\pb=\sum_\pm p_\pm (\xx\pm\yy)/\sqrt{2}+p_z\zz$, where
\begin{subequations} \label{dipolepmz}
\begin{align}
    p_\pm&=\pb\cdot(\xx\mp\ii\yy)/\sqrt{2}, \\
    p_z&=\pb\cdot\zz,
\end{align}
\end{subequations}
Eq.\ \eqref{Edbis} can be rewritten as (see Sec.\ \ref{SIbig})
\begin{align} \nonumber
    \Eb^{\rm dip}(\Rb,\Rb_0) = \pi k^2\sum_m J_m(k_1 R_0) \ee^{-\ii m \varphi_0} \left[ 
    \sum_\pm \frac{p_\pm}{\sqrt{2}}\left(\Eb^H_{1,q (m+1)s}\pm\frac{q}{k_1}\Eb^H_{1,q (m+1)p}\right) +
    \ii   p_z            \frac{Q_1}{k_1}\Eb^H_{1,qmp} \right],
\end{align}
where the fields $\Eb^H_{1,qm\sigma}(\Rb)$ are defined in Eqs.\ \eqref{EqH:Es_Ep}.

% =========================================================
\subsubsection{Normal emission into the far field}
\label{eitff}

The transmission of electromagnetic fields from the waveguide is determined from Eqs.\ \eqref{Eq:Scattering} and \eqref{Eq:RT_coefficients}, which show that, for the special case of $q=0$ considered here, polarization states do not mix (i.e., $t_{m,\sigma\sigma'=0}$ for $\sigma\neq\sigma'$). We thus express the field outside the fiber produced by a line dipole $\pb$ placed at $\Rb_0$ as
\begin{align} \label{Eq:E_out}
    \Eb^{\rm out}(\Rb,\Rb_0) = &\pi k^2\sum_{m} J_m(k_1 R_0) \ee^{-\ii m \varphi_0} \\
    &\times\left[ 
    \frac{p_+}{\sqrt{2}} t_{m+1,ss}\Eb^H_{{\rm h},0 (m+1)s} +
    \frac{p_-}{\sqrt{2}} t_{m-1,ss}\Eb^H_{{\rm h},0 (m-1)s} +
    \ii   p_z            t_{m  ,pp}\Eb^H_{{\rm h},0mp} \right], \nonumber
\end{align}
which is obviously independent of $z$. In the far field, we can use the asymptotic limit $H_m^{(1)}(\theta)\approx \sqrt{2/\pi \theta}\;\ee^{\ii \left(\theta-m\pi/2-\pi/4\right)}$ (see Eq.\ 10.17.5 in Ref.\ \cite{DLMF}) for large arguments of the Hankel functions in the outgoing waves $\Eb^H_{j,0m\sigma}$ (see Eqs.\ \eqref{EqH:Es_Ep}), which allows us to write the electric field as
\begin{align} \nonumber
\Eb^H_{{\rm h},0m \sigma} (\Rb) \xrightarrow[\kh R\gg1]{} \frac{\ee^{\ii \kh R}}{\sqrt{\kh R}}\;\sqrt{\frac{2}{\pi}}\;\ee^{\ii \left(m \varphi -m \pi/2 -\pi/4 \right)}\times
\left\{ \begin{array}{c} -\ii\hat{\varphi}, \quad\quad  (\sigma=s), \\ \;\;\zz, \quad\quad \;\;(\sigma=p), \end{array} \right.
\end{align}
while the magnetic far field is obtained by using Eq.\ \eqref{Eq:s_p_transformation} and Faraday's law as $\Hb^H_{{\rm h},0m \sigma}\xrightarrow[\kh R\gg1]{}-\ii\sqrt{\epsh}\,\Eb^H_{{\rm h},0m \sigma'}$ with $\sigma'\neq\sigma$. Applying these expressions to Eq.\ \eqref{Eq:E_out}, we find the far electric field
\begin{align} \label{Eoutlimit}
    \Eb^{\rm out}(\Rb,\Rb_0) \xrightarrow[\kh R\gg1]{} 
    \left[
     \mathcal{S}^+(\RR,\Rb_0,q,\omega) p_+
    +\mathcal{S}^-(\RR,\Rb_0,q,\omega) p_-
    +\mathcal{S}^z(\RR,\Rb_0,q,\omega) p_z 
    \right ]\frac{\ee^{\ii \kh R}}{\sqrt{\kh R}},
\end{align}
where
\begin{subequations}  \label{Eq:Green_tensor_system}
\begin{align}
\mathcal{S}^\pm(\RR,\Rb_0,\omega) &= \pm\ee^{3\ii\pi/4}\sqrt{\pi} k^2\; \hat{\varphi} \, \sum_{m} \ii^{-m} J_m(k_1 R_0)\;\ee^{\ii [(m\pm1)\varphi-m \varphi_0]}\; t_{m\pm1,ss},\\
\mathcal{S}^z  (\RR,\Rb_0,\omega) &= \ee^{\ii\pi/4}\sqrt{2\pi} k^2\; \zz \, \sum_{m} \ii^{-m} J_m(k_1 R_0) \;\ee^{\ii m(\varphi-\varphi_0)}\; t_{m  ,pp},
\end{align}
\end{subequations}
and the dipole components are defined in Eqs.\ \eqref{dipolepmz}.

The above relations allow us to obtain explicit expressions for the far-field limit ($\kh R\gg1$) of the two-dimensional electromagnetic Green tensor $\Gg_{\rm 2D}(\Rb,\Rb_0,\omega)$, which is implicitly defined through the expression $\Eb^{\rm out}(\Rb,\Rb_0) = \Gg_{\rm 2D}(\Rb,\Rb_0,\omega) \cdot \pb$, relating the strength $\pb$ of a uniform line dipole placed at $\Rb_0$ inside the waveguide to the electric field $\Eb^{\rm out}(\Rb,\Rb_0)$ that it generates at a position $\Rb$ outside it. Taking into consideration the general asymptotic relation $\Gg_{\rm 2D}(\Rb,\Rb_0,\omega)\xrightarrow[\kh R\gg1]{}\left(\ee^{\ii \kh R}/\sqrt{\kh R}\right)\,\Sb(\RR,\Rb_0,\omega)$, and comparing it to Eqs.\ \eqref{Eoutlimit} and \eqref{Eq:Green_tensor_system}, we can readily write $\mathcal{S}^\pm=\Sb\cdot(\xx\pm\ii\yy)/\sqrt{2}$ and $\mathcal{S}^z=\Sb\cdot\zz$ and obtain the explicit formula in Eq.\ \eqref{gexplicit}. These results can be easily generalized to off-normal emission ($q\neq0$), involving off-diagonal transmission coefficients that lead to more involved expressions.

% =========================================================
\subsection{Sum-frequency generation by counter-propagating waveguided pulses}
\label{Sec4.3}

We now introduce counter-propagating pulse fields $\Eb_i(\rb,t)$ and $\Eb_{i'}(\rb,t)$ of the form given in Eq.\ \eqref{Eq:E_Rt_i}, oscillating at frequencies $\omega_{i}$ and $\omega_{i'}$, respectively. Through the second-order nonlinearity of the waveguide material $\chi^{(2)}_{abc}$, where the subscripts $a$, $b$, and $c$ denote Cartesian components, a polarization density $\Pb_{ii'}$ is produced at frequency $\omega_{ii'} = \omega_i+\omega_{i}$. More precisely,
\begin{align} \label{Pfirst}
    \Pb_{ii'}(\rb,t) = |\bar\chi^{(2)}| \tilde{\Pb}_{ii'}(\rb,t),
\end{align}
where $|\bar\chi^{(2)}|$, defined in Eq.\ \eqref{Eq:chi2_average}, is introduced to quantify the strength of the SF susceptibility, while the normalized polarization density can be separated as
\begin{align} \label{Eq:P_a_aux}
    \tilde{\Pb}_{ii'}(\rb,t) = \tilde{\Pb}_{ii'}(\Rb) \; S_{ii'}(z,t) + {\rm c.c.}
\end{align}
by defining $\tilde{\Pb}_{ii'}(\Rb)$ as in Eq.\ \eqref{Eq:P_a_auxbis}, as well as the $(z,t)$-dependent factor
\begin{align} \label{Siit}
    S_{ii'}(z,t) = \ee^{\ii [(q_i+q_{i'})z-\omega_{ii'}t]} \ee^{-(z-v_i t)^2/L^2} \ee^{-(z-v_{i'} t)^2/L^2}.
\end{align}
Eventually, we set $q_i=-q_{i'}$ (i.e., waveguided photons with opposite wave vectors, leading to normal emission of SF photons), so $v_i$ and $v_{i'}$ also have opposite signs inherited from $q_i$ and $q_{i'}$. In addition, the nonlinear susceptibility $\chi^{(2)}_{abc}(\rb)$ is taken to be constant over the range of frequencies under consideration, as well as uniform inside the fiber and zero outside it. We note that the present analysis could be trivially extended to consider a nonlinear cladding instead.

The SF field $\Eb_{ii'}(\rb,t)$ is thus produced by the nonlinear polarization density according to
\begin{align} \label{Eii1}
    \Eb_{ii'}(\rb,t) = \int \frac{d\omega}{2\pi}\, \ee^{-\ii \omega t} \, \Eb_{ii'}(\rb,\omega) = \int \frac{d\omega}{2\pi}  \ee^{-\ii \omega t} \int d^3\rb' \, \Gg(\rb,\rb',\omega) \cdot \int dt' \ee^{\ii\omega t'} \Pb_{ii'}(\rb',t'),
\end{align}
where we have introduced the three-dimensional electromagnetic Green tensor $\Gg(\rb,\rb',\omega)$, defined in such a way that $\Eb(\rb,\rb') = \Gg(\Rb,\rb',\omega) \cdot \pb$ gives the field produced at $\rb$ by a point dipole of strength $\pb$ placed at $\rb'$. To quantify SF generation in the far-field limit (i.e., at $\kh r\gg1$), we exploit the translational invariance of the polarization source to write
\begin{align} \label{Eq:Green_tensor_approx} 
    \Gg(\rb,\rb',\omega) = \Gg(\rb-z'\zz,\Rb',\omega) \xrightarrow[\kh r\gg1]{} \frac{\ee^{\ii \kh r }}{r} \ee^{-\ii \kh zz'/r}\gb(\rr,\Rb',\omega),
\end{align}
which allows us to recast Eq.\ \eqref{Eii1} as
\begin{align} \label{Eq:E_SF_FF}
    \Eb_{ii'}(\rb,t) = \int \frac{d\omega}{2\pi} \,\ee^{-\ii \omega t}\, \frac{\ee^{\ii \kh r }}{r} \fb_{ii'}(\rr,\omega)
\end{align}
in terms of the frequency-space electric far-field amplitude
\begin{align} \label{fiiw}
    \fb_{ii'}(\rr,\omega) = \int d^3\rb' \, \ee^{-\ii \kh zz'/r}\gb(\rr,\Rb',\omega) \cdot\int dt' \ee^{\ii\omega t'}\Pb_{ii'}(\rb',t').
\end{align}
Finally, as shown in Sec.\ \ref{gtoS}, we can relate the far-field three-dimensional amplitude obtained from Green tensor in Eqs.\ \eqref{Eq:Green_tensor_approx} and \eqref{fiiw} to the two-dimensional one presented in Eqs.\ \eqref{Eq:Green_tensor_system} as
\begin{align} \nonumber
    \gb^j(\rr,\Rb',\omega) = \frac{\ee^{-\ii \pi/4}}{\sqrt{2\pi}}\Sb(\RR,\Rb',q,\omega),
\end{align}
with explicit expressions for the components of $\Sb(\RR,\Rb',q,\omega)$ also offered in Sec.\ \ref{eitff}.

% =========================================================
\subsection{Up- and down-conversion efficiency}
\label{updown}

The efficiency of the SPDC process in which a photon impinging normally to the waveguide direction produces a pair of guided photons moving in opposite directions away from one another is argued to be identical with that of an up-conversion process involving SF generation of two guided photons moving towards one another, provided that reciprocity applies. We then calculate the SF efficiency $\eta_{ii'}=N_{ii'}/N_iN_{i'}$ as the ratio of the number of emitted SF photons $N_{ii'}$ to the number of incident photons $N_i$ and $N_{i'}$ in both guided pulses.

The number of photons carried by a guided mode pulse with field profile $\Eb_i(\rb,t)$ is expressed as
\begin{align} \nonumber
    N_i = \frac{1}{\hbar\omega_i} \int_{-\infty}^{\infty} dt \int d^2\Rb\,  {\bf S}_i(\rb,t) \cdot \zz,
\end{align}
where we evaluate the flux carried by the Poynting vector ${\bf S }_i(\rb,t) = (c/4\pi) \Eb_i(\rb,t) \times \Hb_i(\rb,t)$ in the $\Rb$ plane. Inserting the fields given by Eqs.\ \eqref{Eq:E_Rt_i} and \eqref{Eq:H_Rt_i} and integrating over time, we obtain
\begin{align} \label{Ni}
N_i = \frac{c L}{2\sqrt{2\pi} |v_i| } \frac{1}{\hbar\omega_i}  \int d^2\Rb \;  {\rm Re}\left\{E_{i,x}(\Rb)H_{i,y}^*(\Rb)-E_{i,y}(\Rb)H^*_{i,x}(\Rb)\right\},
\end{align}
Likewise, the number of photons produced in the far field via SF generation from the radial component of the energy emanating from the fiber can be in turn obtained from the far-field Poynting vector $\Sb^\infty_{ii'}$ (evaluated from the field in Eq.\ \eqref{Eq:E_SF_FF}) as
\begin{align}
    N_{ii'} &= \frac{1}{\hbar\omega_{ii'}} \int_{-\infty}^{\infty} dt \int d\Omega_{\rr} \, r^2 \, \RR\cdot\Sb^\infty_{ii'}(\rb,t) = \frac{\sqrt{\eps_{\rm h}}c}{4\pi^2\hbar\omega_{ii'}} \int d\Omega_{\rr} \, 
    (\RR\cdot\rr) \int_0^\infty d\omega
    \left|\fb_{ii'}(\rr,\omega)\right|^2. \label{Niifirst}
\end{align}
In the derivation of this expression, we have used the fact that $\fb_{ii'}(\rr,\omega)\cdot\rr=0$ (i.e., the far field is transverse). For long pulses (see Eq.\ \eqref{Eq:E_Rt_i}), only the first term in Eq.\ \eqref{Eq:P_a_aux} (peaked around frequencies $\omega\sim\omega_{ii'}$) contributes to $\fb_{ii'}(\rr,\omega)$ over the $\omega>0$ integral in Eq.\ \eqref{Niifirst}, and therefore, using Eqs.\ \eqref{Pfirst}, \eqref{Eq:P_a_aux}, and \eqref{fiiw}, we can write
\begin{align} \nonumber
    \int_0^\infty d\omega \left|\fb_{ii'}(\rr,\omega)\right|^2 \approx 2\pi|\bar\chi^{(2)}|^2 A_{ii'}\left|\;\int_{R'<a}\!\!\! d^2\Rb' \, \gb(\rr,\Rb',\omega_{ii'}) \cdot\tilde\Pb_{ii'}(\Rb')\right|^2.
\end{align}
where
\begin{align} 
   A_{ii'}
   &=\frac{1}{2\pi}\int_0^\infty d\omega \left|\int dz' \int dt \,\ee^{-\ii \kh zz'/r}\,\ee^{\ii\omega t}\,S_{ii'}(z',t)\right|^2 \nonumber\\
   &\approx\frac{1}{2\pi}\int_{-\infty}^\infty d\omega \left|\int dz' \int dt \,\ee^{-\ii \kh zz'/r}\,\ee^{\ii\omega t}\,S_{ii'}(z',t)\right|^2 \nonumber\\
   &=\int dt \left|\int dz' \,\ee^{-\ii \kh zz'/r}\,S_{ii'}(z',t)\right|^2, \nonumber\\
   &=\frac{\pi^{3/2} L^3}{2\vert v_{i}-v_{i'} \vert}\ee^{-(q_i+q_{i'}-q)^2L^2/4} \nonumber
\end{align}
with $q=(\sqrt{\eps_{\rm h}}\omega_{ii'}/c) z/r$. Here, we have approximated $\omega$ by $\omega_{ii'}$ in $\gb(\rr,\Rb',\omega)$ and $\kh$ (because the incident pulses are long enough to be considered narrowly peaked around their respective central frequencies $\omega_i$ and $\omega_{i'}$), extended the $\omega$ range of integration to nonresonant negative values, used the Percival theorem, and taken $S_{ii'}(z',t)$ from Eq.\ \eqref{Siit} to analytically evaluate the remaining integrals. In addition, we can carry out the polar-angle integral in Eq.\ \eqref{Niifirst} by writing $d\Omega_{\rr}=d\varphi dq\times(c/\sqrt{\eps_{\rm h}}\omega_{ii'})$ and considering that the pulse length $L$ is sufficiently large as to make $A_{ii'}$ negligible outside the light cone $|q|<\sqrt{\eps_{\rm h}}\omega_{ii'}/c$. This leads to
\begin{align} \label{Nphi}
    N_{ii'} = \int_0^{2\pi}\; d\varphi N_{ii'}(\varphi)
\end{align}
with
\begin{align} \label{Nii}
    N_{ii'}(\varphi) =|\bar\chi^{(2)}|^2 \frac{L^2 c^2}{4\hbar\omega_{ii'}^2\vert v_{i}-v_{i'} \vert}
    (\RR\cdot\rr) \left|\int_{R'<a}\!\!\! d^2\Rb' \, \gb(\rr,\Rb',\omega_{ii'}) \cdot\tilde\Pb_{ii'}(\Rb')\right|^2,
\end{align}
where the direction $\rr$ is defined by the azimuthal angle $\varphi$ and the component $q_i+q_{i'}$ of the emitted light wave vector along the fiber axis.

Finally, we specialize the above expressions to $q_i+q_{i'}=0$ (normal emission, for which $\RR\cdot\rr=1$) and evaluate the efficiency $\eta_{ii'}=N_{ii'}/N_iN_{i'}$ by using Eqs.\ \eqref{Ni}, \eqref{Nphi}, and \eqref{Nii}, from which we find the result shown in Eqs.\ \eqref{eta} and \eqref{etafinal}, where we have explicitly indicated the dependence of the far-field Green tensor
\begin{align}
\label{gexplicit}
\gb(\rr,\Rb',\omega)&=\gb(\varphi-\varphi',R',\omega) \\
&=k^2\; \sum_{m} \ii^{-m} J_m(k_1 R')\;\ee^{\ii m(\varphi-\varphi')}\;
\left[t_{m,pp} \, \zz\otimes\zz
+\frac{1}{2}\sum_{\pm} t_{m\pm1,ss}\, \hat{\varphi}\otimes(\hat{\varphi}\pm\ii\RR) \right] \nonumber
\end{align}
on the difference of the azimuthal angles of $\rr$ and $\Rb'$ in the $\{\RR,\hat\varphi,\zz\}$ frame (see Sec.\ \ref{gtoS}).

% =========================================================
% --- reflection and transmission coefficients ------------
% =========================================================
\section{Reflection and transmission coefficients of inner cylindrical waves at the waveguide interface}
\label{tpts}

Following the notation introduced in Sec.\ \ref{Sec4.1} of the main text, we first note that $m$ and $q$ are unchanged upon reflection or transmission due to the cylindrical symmetry of the waveguide. The corresponding coefficients $r_{m,\sigma'\sigma}$ and $t_{m,\sigma'\sigma}$ for a cylindrical wave of electric field $\Eb^H_{1,q m \sigma}$ and polarization $\sigma\in\{s,p\}$ emanating from inside the waveguide are defined through the expression
\begin{align}
    \Eb = \begin{cases} \Eb^H_{1,q m \sigma}+r_{m,s\sigma} \Eb^J_{1,qms} +  r_{m,p\sigma} \Eb^J_{1,qmp},&\ R<a, \\ 
    t_{m,s\sigma}\Eb^H_{h,qms} + t_{m,p\sigma} \Eb^H_{h,qmp},&\ R\geq a, \end{cases} \nonumber
\end{align}
where the reflected waves $\Eb^J_{1,qm\sigma'}$ are regular propagating solutions inside the waveguide material $j=1$, while the transmitted fields $\Eb^H_{h,qm\sigma'}$ are outgoing solutions in the host medium $j=h$. We now enforce the continuity of the tangential components (i.e., perpendicular to $\RR$) in both magnetic and electric fields, where the former is obtained from the latter by using Faraday's law ($\Hb^{J/H}_{j,qm\sigma} = (1/\ii k)\nabla \times \Eb^{J/H}_{j,qm\sigma}$) combined with the identity \cite{paper047} $k_j\Eb^{J/H}_{j,qm\sigma} = \nabla \times \Eb^{J/H}_{j,qm\sigma'}$, valid for $\sigma\neq\sigma'$. This leads to the expressions
\begin{subequations} \label{system}
\begin{align}
\left[r_{m,s\sigma} \Eb^J_{1,qms}-t_{m,s \sigma}\Eb^H_{h,qms} +  r_{m,p\sigma} \Eb^J_{1,qmp} - t_{m,p\sigma} \Eb^H_{h,qmp}\right]\times\RR &= \;\;-\Eb^H_{1,q m \sigma}\times\RR, \\
\left[\zeta r_{m,s\sigma} \Eb^J_{1,qmp}-t_{m,s \sigma}\Eb^H_{h,qmp} +\zeta r_{m,p\sigma} \Eb^J_{1,qms} - t_{m,p\sigma} \Eb^H_{h,qms}\right]\times\RR &= -\zeta \Eb^H_{1,q m \sigma'}\times\RR,
\end{align}
\end{subequations}
where $\zeta = \sqrt{\eps_1 / \eps_h}$ and $\sigma \neq \sigma'$. By inserting the explicit expressions of the mode fields into Eqs.\ \eqref{system} and projecting on $\hat{\varphi}$ and $\zz$ components, we readily find the secular matrix $M$ and the linear equations for the reflection and transmission coefficients given in Sec.\ \ref{Sec4.1}. Waveguide modes are signalled by the zeros of $\det\{M\}$, as discussed in Sec.\ \ref{Sec4.1.1} of the main text. We show the dispersion relation and group velocity of the lowest-order modes for $\epsc=5$ and $\epsh=1$ in Fig.\ \ref{FigS1}.

% =========================================================
% --- plane wave in cylindrical waves ---------------------
% =========================================================
\section{Decomposition of a light plane wave in cylindrical waves} \label{planewave}
\label{planetocyl}

We work in frequency space $\omega$ and consider a light plane wave propagating in a medium of permittivity $\eps_1$ with unit electric field $\eh^\pm_\sigma\ee^{\ii\kb_1^{\pm}\cdot\rb}$ of polarization $\sigma\in\{s,p\}$ and wave vector $\kb^\pm_1=\Qb\pm k_{1z}\zz$. Here, $\Qb=(Q_x,Q_y)=(Q,\varphi_\Qb)$, $k_{1z}=\sqrt{\eps_1k^2-Q^2+\ii0^+}$ (with $k=\omega/c$ and ${\rm Re}\{k_{1z}\}>0$), and the polarization vectors are defined as $\eh^\pm_s=(-Q_y\xx+Q_x\yy)/Q$ and $\eh^\pm_p=(\pm\Qb k_z-Q^2\zz)/k_1Q$. It is convenient to recast the dependence on $\Rb=(x,y)=(R,\varphi)$ by using the orthogonality relation $\int_0^\infty RdR\,J_m(QR)J_m(Q'R)=\delta(Q-Q')/Q$ together with the integral $\int_0^{2\pi}d\varphi\,\ee^{\ii\Qb\cdot\Rb}\ee^{\ii m\varphi}=2\pi\ii^mJ_m(QR)\ee^{\ii m\varphi_\Qb}$, from which we derive the Fourier expansion $\ee^{\ii\Qb\cdot\Rb}=\sum_m\ii^mJ_m(QR)\,\ee^{\ii m(\varphi-\varphi_\Qb)}$. Combining this result together with the explicit forms of the polarization vectors given above, we can assimilate each of the $m$ terms in the Fourier transform of the plane wave field to a cylindrical wave and write
\begin{align}
&\eh^\pm_s\ee^{\ii\kb_1^{\pm}\cdot\rb}=\sum_m\ii^{m+1}\,\ee^{-\ii m\varphi_\Qb}\,\Eb^J_{1,\pm k_{1z}ms}, \nonumber \\
&\eh^\pm_p\ee^{\ii\kb_1^{\pm}\cdot\rb}=-\sum_m\ii^m\,\ee^{-\ii m\varphi_\Qb}\,\Eb^J_{1,\pm k_{1z}mp}, \nonumber
\end{align}
where $\Eb^J_{1,qm\sigma}$ is defined in Sec.\ \ref{Sec4.1} in the main text.

% =========================================================
% --- details of emission ---------------------------------
% =========================================================
\section{Decomposition of the field due to a line dipole in cylindrical waves}
\label{SIbig}

In Sec.\ \ref{Sec4.2} of the main text, we express the field produced by a line dipole $\pb$ in a homogeneous medium $\eps_1$ as
\begin{align} \nonumber
    \Eb^{\rm dip}(\rb,\Rb_0) =
    \frac{\ii \pi}{\eps_1} \sum_m J_m(Q_1 R_0) \ee^{-\ii m \varphi_0} \left[k_1^2\pb + \nabla (\pb \cdot \nabla) \right] \, H_m^{(1)}(Q_1 R) \, \ee^{\ii m \varphi}\ee^{\ii qz},
\end{align}
where the $\ee^{\ii qz}$ dependence is inherited from the modulation of the line dipole along $z$. We can express this field in terms of cylindrical waves by projecting the dipole on the circular coordinate vectors $\eh^\pm=(\xx\pm\ii\yy)/\sqrt{2}=\ee^{\pm\ii\varphi}(\RR\pm\ii\hat\varphi)/\sqrt{2}$, such that
\begin{align} \nonumber
    \pb = p_+\eh^++p_-\eh^-+p_z\zz
\end{align}
with coordinates
\begin{align}
    p_\pm&=\pb\cdot\eh^\mp=\pb\cdot(\xx\mp\ii\yy)/\sqrt{2}, \nonumber\\
    p_z&=\pb\cdot\zz. \nonumber
\end{align}
Working out the $p_z$ contribution to $\Eb^{\rm dip}$ and comparing it to the cylindrical waves in Sec.\ \ref{Sec4.1}, we find that it reduces to $p_zQ_1k_1\Eb^H_{1,qmp}$. The $p_\pm$ contributions are more involved, as they contain both $p$ and $s$ waves. More precisely, the $\Eb^{\rm dip}$ field that they generate has a $\zz$ component, which can be assigned to a $p$ cylindrical wave $\propto\Eb^H_{1,q(m\pm1)p}$. Adding and subtracting this wave to eliminate the $z$ component, we find that the remaining $\RR$ and $\hat\varphi$ components reduce to a wave $\propto\Eb^H_{1,q(m\pm1)s}$ of $s$ polarization. Combining these results, we can write the dipole field as
\begin{align} \nonumber
    \Eb^{\rm dip}(\Rb,\Rb_0) = \pi k^2\sum_m J_m(k_1 R_0) \ee^{-\ii m \varphi_0} \left[ 
    \sum_\pm \frac{p_\pm}{\sqrt{2}}\left(\Eb^H_{1,q (m+1)s}\pm\frac{q}{k_1}\Eb^H_{1,q (m+1)p}\right) +
    \ii   p_z            \frac{Q_1}{k_1}\Eb^H_{1,qmp} \right],
\end{align}
which is the expression reproduced in Sec.\ \ref{Sec4.2}. In the algebraic manipulations needed to carry out these derivations, we make intensive use of the relations
\begin{align}
&\frac{m}{\theta}\mathcal{C}_m=\frac{1}{2}(\mathcal{C}_{m-1}+\mathcal{C}_{m+1}), \nonumber \\
&\mathcal{C}'_m=\frac{1}{2}(\mathcal{C}_{m-1}-\mathcal{C}_{m+1}), \nonumber \\
&\mathcal{C}''_m=-\mathcal{C}_m+\frac{m-1}{2\theta}\mathcal{C}_{m-1}+\frac{m+1}{2\theta}\mathcal{C}_{m+1} \nonumber
\end{align}
for the Bessel and Hankel functions $\mathcal{C}_m(\theta)$, which can be directly obtained from the recurrence relation $\mathcal{C}'_m(\theta)=\pm(m/\theta)\mathcal{C}_m(\theta)\mp\mathcal{C}_{m\pm1}(\theta)$ \cite{DLMF}.

% =========================================================
% --- far-field Green tensor ------------------------------
% =========================================================
\section{Far-field limit of the electromagnetic Green tensor}
\label{gtoS}

Because the waveguide is translationally invariant along $z$, the Green tensor satisfies the identity
\begin{align} \label{limit1}
\Gg(\rb,\rb',\omega) = \Gg(\rb-z'\zz,\Rb',\omega) \xrightarrow[\kh r\gg1]{} \frac{\ee^{\ii \kh r }}{r} \ee^{-\ii \kh zz'/r}\gb(\rr,\Rb',\omega),
\end{align}
where the rightmost expression represents the far-field limit in the host medium, for which we implicitly define a tensor $\gb(\rr,\Rb',\omega)$ that depends only on the direction of $\rb$. In the derivation of this result, we have approximated $|\rb-z'\zz|\approx r-zz'/r$ in the leading exponential, assuming that we have $r\gg r'$. Translational symmetry also allows us to represent the Green tensor in wave vector space along $z$ and $z'$ according to
\begin{align} \nonumber
    \Gg(\rb,\rb',\omega) = \int \frac{dq}{2\pi}\, \Gg_{\rm 2D}(\Rb,\Rb',q,\omega)\, \ee^{\ii q(z-z')},
\end{align}
where the two-dimensional Green tensor $\Gg_{\rm 2D}(\Rb,\Rb',q,\omega)$ has the far-field behavior
\begin{align} \nonumber
\Gg_{\rm 2D}(\Rb,\Rb',q,\omega) \xrightarrow[\kh R\gg1]{} \frac{\ee^{\ii \Qh R}}{\sqrt{\Qh R}} \Sb(\RR,\Rb',q,\omega)
\end{align}
with $\Qh=\sqrt{k_{\rm h}^2-q^2+\ii0^+}$ defined in the same way as in Sec.\ \ref{Sec4.1} of the main text. Combining the above expressions, we have
\begin{align} \nonumber
    \Gg(\rb,\rb',\omega) \xrightarrow[\kh R\gg1]{} \int \frac{dq}{2\pi}\, \frac{\ee^{\ii(\Qh R+\ii qz)}}{\sqrt{\Qh R}} \ee^{-\ii qz'} \Sb(\RR,\Rb',q,\omega).
\end{align}
We now work out the $q$ integral in the asymptotic limit by following the stationary-phase method. More precisely, we approximate $\Qh R+qz\approx\kh r-(k_{\rm h}^2R/2Q_0^3)(q-q_0)^2$ by its second-order Taylor expansion around the stationary point defined by the vanishing of its first derivative $-q_0R/Q_0+z=0$ with $Q_0=\sqrt{k_{\rm h}^2-q_0^2}$ (i.e., with $q_0<\kh$ such that $(Q_0,q_0)\parallel(R,z)$, and therefore, $Q_0=\kh R/r$ and $q_0=\kh z/r$). Since only the region very close to $q_0$ contributes to the integral in the far-field limit, we can set  $q=q_0$ in the rest of the integrand and write
\begin{align} \label{limit2}
    \Gg(\rb,\rb',\omega) \xrightarrow[\kh R\gg1]{} \frac{\ee^{\ii\kh r}}{2\pi\sqrt{\kh R^2/r}}\ee^{-\ii\kh zz'/r} \Sb(\RR,\Rb',q_0,\omega)\int dq\,\ee^{-\ii q^2r^3/2\kh R^2}=\frac{\ee^{\ii\kh r}}{r}\ee^{-\ii\kh zz'/r} \frac{\ee^{-\ii\pi/4}}{\sqrt{2\pi}}\Sb(\RR,\Rb',q_0,\omega),
\end{align}
where the right-most expression is obtained by applying the integral $\int_{-\infty}^{\infty} d\theta\,\ee^{\ii \theta^2}=\sqrt{\pi}\,\ee^{-\ii\pi/4}$. Comparing Eqs.\ \eqref{limit1} and  \eqref{limit2}, we find
\begin{align} \label{verylast}
\gb(\rr,\Rb',\omega)=\frac{\ee^{-\ii\pi/4}}{\sqrt{2\pi}}\Sb(\RR,\Rb',q_0,\omega),
\end{align}
where $q_0=\kh z/r$. We use this relation in the main text to find an explicit expression for $\gb(\rr,\Rb',\omega)$ based on the far-field limit of the outgoing cylindrical waves generated by a line dipole placed inside the waveguide for the particular case of $q=0$ (normal emission) with $\Sb(\RR,\Rb',\omega)\equiv\Sb(\RR,\Rb',q=0,\omega)$. Gathering the results in Secs.\ \ref{Sec4.2} and \ref{Sec4.3}, together with Eq.\ \eqref{verylast}, we obtain
\begin{align}
\gb(\rr,\Rb',\omega)=
&k^2\; \sum_{m} \ii^{-m} J_m(k_1 R')\;\ee^{\ii m(\varphi-\varphi')}\;
\left[t_{m,pp} \, \zz\otimes\zz
+\frac{1}{2}\sum_{\pm} t_{m\pm1,ss}\, \hat{\varphi}\otimes(\hat{\varphi}\pm\ii\RR) \right], \nonumber
\end{align}
whose components in the $\{\RR,\hat\varphi,\zz\}$ frame depend on $\varphi$ and $\varphi'$ only through the difference $\varphi-\varphi'$, thus reflecting the cylindrical symmetry of the system.

% =========================================================
% --- SI, acknowledgments, bibliography, etc. -------------
% =========================================================

% --- SI --------------------------------------------------
%\section*{Supplementary Information} %---ACS optional
%The Supporting Information is available free of charge at https://pubs.acs.org/doi/xxx. %---ACS optional

% --- acknowledgments -------------------------------------
%\section*{Funding} % ... text ... %---OSA, place funding here and thanks to colleagues below
\section*{ACKNOWLEDGEMENTS} %---ACS---arxiv
%\section*{Acknowledgments} %---OSA
%\begin{acknowledgments} %---APS
{\bf Funding:} This work has been supported in part by ERC (Advanced Grant 789104-eNANO), the Spanish MINECO (PID2020-112625GB-I00 and SEV2015-0522), the Catalan CERCA Program, the Generalitat de Catalunya, the European Social Fund (L\text{'}FSE inverteix en el teu futur)-FEDER. J.~D.~C. is a Sapere Aude research leader supported by Independent Research Fund Denmark (grant no. 0165-00051B). The Center for Nano Optics is financially supported by the University of Southern Denmark (SDU 2020 funding).

\end{widetext} %---arxiv

%\bibliographystyle{apsrev}
%\bibliography{../../../bibtex/refsL.bib} 

\begin{thebibliography}{41}
\expandafter\ifx\csname natexlab\endcsname\relax\def\natexlab#1{#1}\fi
\expandafter\ifx\csname bibnamefont\endcsname\relax
  \def\bibnamefont#1{#1}\fi
\expandafter\ifx\csname bibfnamefont\endcsname\relax
  \def\bibfnamefont#1{#1}\fi
\expandafter\ifx\csname citenamefont\endcsname\relax
  \def\citenamefont#1{#1}\fi
\expandafter\ifx\csname url\endcsname\relax
  \def\url#1{\texttt{#1}}\fi
\expandafter\ifx\csname urlprefix\endcsname\relax\def\urlprefix{URL }\fi
\providecommand{\bibinfo}[2]{#2}
\providecommand{\eprint}[2][]{\url{#2}}

\bibitem[{\citenamefont{Zheng and Guo}(2000)}]{ZG00}
\bibinfo{author}{\bibfnamefont{S.-B.} \bibnamefont{Zheng}} \bibnamefont{and}
  \bibinfo{author}{\bibfnamefont{G.-C.} \bibnamefont{Guo}},
  \bibinfo{journal}{Phys.\ Rev.\ Lett.} \textbf{\bibinfo{volume}{85}},
  \bibinfo{pages}{2392} (\bibinfo{year}{2000}).

\bibitem[{\citenamefont{Raimond et~al.}(2001)\citenamefont{Raimond, Brune, and
  Haroche}}]{RBH01}
\bibinfo{author}{\bibfnamefont{J.-M.} \bibnamefont{Raimond}},
  \bibinfo{author}{\bibfnamefont{M.}~\bibnamefont{Brune}}, \bibnamefont{and}
  \bibinfo{author}{\bibfnamefont{S.}~\bibnamefont{Haroche}},
  \bibinfo{journal}{Rev.\ Mod.\ Phys.} \textbf{\bibinfo{volume}{73}},
  \bibinfo{pages}{565} (\bibinfo{year}{2001}).

\bibitem[{\citenamefont{Lamas-Linares et~al.}(2001)\citenamefont{Lamas-Linares,
  Howell, and Bouwmeester}}]{LHB01}
\bibinfo{author}{\bibfnamefont{A.}~\bibnamefont{Lamas-Linares}},
  \bibinfo{author}{\bibfnamefont{J.~C.} \bibnamefont{Howell}},
  \bibnamefont{and}
  \bibinfo{author}{\bibfnamefont{D.}~\bibnamefont{Bouwmeester}},
  \bibinfo{journal}{Nature} \textbf{\bibinfo{volume}{412}},
  \bibinfo{pages}{887} (\bibinfo{year}{2001}).

\bibitem[{\citenamefont{Monroe}(2002)}]{M02_1}
\bibinfo{author}{\bibfnamefont{C.}~\bibnamefont{Monroe}},
  \bibinfo{journal}{Nature} \textbf{\bibinfo{volume}{416}},
  \bibinfo{pages}{238} (\bibinfo{year}{2002}).

\bibitem[{\citenamefont{Inagaki et~al.}(2013)\citenamefont{Inagaki, Matsuda,
  Tadanaga, Asobe, and Takesue}}]{IMT13}
\bibinfo{author}{\bibfnamefont{T.}~\bibnamefont{Inagaki}},
  \bibinfo{author}{\bibfnamefont{N.}~\bibnamefont{Matsuda}},
  \bibinfo{author}{\bibfnamefont{O.}~\bibnamefont{Tadanaga}},
  \bibinfo{author}{\bibfnamefont{M.}~\bibnamefont{Asobe}}, \bibnamefont{and}
  \bibinfo{author}{\bibfnamefont{H.}~\bibnamefont{Takesue}},
  \bibinfo{journal}{Opt.\ Express} \textbf{\bibinfo{volume}{21}},
  \bibinfo{pages}{23241} (\bibinfo{year}{2013}).

\bibitem[{\citenamefont{Liao et~al.}(2017)\citenamefont{Liao, Cai, Liu, Zhang,
  Li, Ren, Yin, Shen, Cao, Li et~al.}}]{LCL17}
\bibinfo{author}{\bibfnamefont{S.-K.} \bibnamefont{Liao}},
  \bibinfo{author}{\bibfnamefont{W.-Q.} \bibnamefont{Cai}},
  \bibinfo{author}{\bibfnamefont{W.-Y.} \bibnamefont{Liu}},
  \bibinfo{author}{\bibfnamefont{L.}~\bibnamefont{Zhang}},
  \bibinfo{author}{\bibfnamefont{Y.}~\bibnamefont{Li}},
  \bibinfo{author}{\bibfnamefont{J.-G.} \bibnamefont{Ren}},
  \bibinfo{author}{\bibfnamefont{J.}~\bibnamefont{Yin}},
  \bibinfo{author}{\bibfnamefont{Q.}~\bibnamefont{Shen}},
  \bibinfo{author}{\bibfnamefont{Y.}~\bibnamefont{Cao}},
  \bibinfo{author}{\bibfnamefont{Z.-P.} \bibnamefont{Li}},
  \bibnamefont{et~al.}, \bibinfo{journal}{Nature}
  \textbf{\bibinfo{volume}{549}}, \bibinfo{pages}{43} (\bibinfo{year}{2017}).

\bibitem[{\citenamefont{Chang et~al.}(2014)\citenamefont{Chang, Vuleti\'{c},
  and Lukin}}]{CVL14}
\bibinfo{author}{\bibfnamefont{D.~E.} \bibnamefont{Chang}},
  \bibinfo{author}{\bibfnamefont{V.}~\bibnamefont{Vuleti\'{c}}},
  \bibnamefont{and} \bibinfo{author}{\bibfnamefont{M.~D.} \bibnamefont{Lukin}},
  \bibinfo{journal}{Nat.\ Photon.} \textbf{\bibinfo{volume}{8}},
  \bibinfo{pages}{685} (\bibinfo{year}{2014}).

\bibitem[{\citenamefont{Li et~al.}(2005)\citenamefont{Li, Voss, Sharping, and
  Kumar}}]{LVS05}
\bibinfo{author}{\bibfnamefont{X.}~\bibnamefont{Li}},
  \bibinfo{author}{\bibfnamefont{P.~L.} \bibnamefont{Voss}},
  \bibinfo{author}{\bibfnamefont{J.~E.} \bibnamefont{Sharping}},
  \bibnamefont{and} \bibinfo{author}{\bibfnamefont{P.}~\bibnamefont{Kumar}},
  \bibinfo{journal}{Phys.\ Rev.\ Lett.} \textbf{\bibinfo{volume}{94}},
  \bibinfo{pages}{053601} (\bibinfo{year}{2005}).

\bibitem[{\citenamefont{Liu et~al.}(2020)\citenamefont{Liu, Nape, Wang,
  Vall{\'e}s, Wang, and Forbes}}]{LNW20}
\bibinfo{author}{\bibfnamefont{J.}~\bibnamefont{Liu}},
  \bibinfo{author}{\bibfnamefont{I.}~\bibnamefont{Nape}},
  \bibinfo{author}{\bibfnamefont{Q.}~\bibnamefont{Wang}},
  \bibinfo{author}{\bibfnamefont{A.}~\bibnamefont{Vall{\'e}s}},
  \bibinfo{author}{\bibfnamefont{J.}~\bibnamefont{Wang}}, \bibnamefont{and}
  \bibinfo{author}{\bibfnamefont{A.}~\bibnamefont{Forbes}},
  \bibinfo{journal}{Sci.\ Adv.} \textbf{\bibinfo{volume}{6}},
  \bibinfo{pages}{eaay0837} (\bibinfo{year}{2020}).

\bibitem[{\citenamefont{Mair et~al.}(2001)\citenamefont{Mair, Vaziri, Weihs,
  and Zeilinger}}]{MVW01}
\bibinfo{author}{\bibfnamefont{A.}~\bibnamefont{Mair}},
  \bibinfo{author}{\bibfnamefont{A.}~\bibnamefont{Vaziri}},
  \bibinfo{author}{\bibfnamefont{G.}~\bibnamefont{Weihs}}, \bibnamefont{and}
  \bibinfo{author}{\bibfnamefont{A.}~\bibnamefont{Zeilinger}},
  \bibinfo{journal}{Nature} \textbf{\bibinfo{volume}{412}},
  \bibinfo{pages}{313} (\bibinfo{year}{2001}).

\bibitem[{\citenamefont{Fickler et~al.}(2012)\citenamefont{Fickler, Lapkiewicz,
  Plick, Krenn, Schaeff, Ramelow, and Zeilinger}}]{FLP12}
\bibinfo{author}{\bibfnamefont{R.}~\bibnamefont{Fickler}},
  \bibinfo{author}{\bibfnamefont{R.}~\bibnamefont{Lapkiewicz}},
  \bibinfo{author}{\bibfnamefont{W.~N.} \bibnamefont{Plick}},
  \bibinfo{author}{\bibfnamefont{M.}~\bibnamefont{Krenn}},
  \bibinfo{author}{\bibfnamefont{C.}~\bibnamefont{Schaeff}},
  \bibinfo{author}{\bibfnamefont{S.}~\bibnamefont{Ramelow}}, \bibnamefont{and}
  \bibinfo{author}{\bibfnamefont{A.}~\bibnamefont{Zeilinger}},
  \bibinfo{journal}{Science} \textbf{\bibinfo{volume}{338}},
  \bibinfo{pages}{640} (\bibinfo{year}{2012}).

\bibitem[{\citenamefont{Malik et~al.}(2016)\citenamefont{Malik, Erhard, Huber,
  Krenn, Fickler, and Zeilinger}}]{MEH16}
\bibinfo{author}{\bibfnamefont{M.}~\bibnamefont{Malik}},
  \bibinfo{author}{\bibfnamefont{M.}~\bibnamefont{Erhard}},
  \bibinfo{author}{\bibfnamefont{M.}~\bibnamefont{Huber}},
  \bibinfo{author}{\bibfnamefont{M.}~\bibnamefont{Krenn}},
  \bibinfo{author}{\bibfnamefont{R.}~\bibnamefont{Fickler}}, \bibnamefont{and}
  \bibinfo{author}{\bibfnamefont{A.}~\bibnamefont{Zeilinger}},
  \bibinfo{journal}{Nat.\ Photon.} \textbf{\bibinfo{volume}{10}},
  \bibinfo{pages}{248} (\bibinfo{year}{2016}).

\bibitem[{\citenamefont{Stav et~al.}(2018)\citenamefont{Stav, Faerman, Maguid,
  Oren, Kleiner, Hasman, and Segev}}]{SFM18}
\bibinfo{author}{\bibfnamefont{T.}~\bibnamefont{Stav}},
  \bibinfo{author}{\bibfnamefont{A.}~\bibnamefont{Faerman}},
  \bibinfo{author}{\bibfnamefont{E.}~\bibnamefont{Maguid}},
  \bibinfo{author}{\bibfnamefont{D.}~\bibnamefont{Oren}},
  \bibinfo{author}{\bibfnamefont{V.}~\bibnamefont{Kleiner}},
  \bibinfo{author}{\bibfnamefont{E.}~\bibnamefont{Hasman}}, \bibnamefont{and}
  \bibinfo{author}{\bibfnamefont{M.}~\bibnamefont{Segev}},
  \bibinfo{journal}{Science} \textbf{\bibinfo{volume}{361}},
  \bibinfo{pages}{1101} (\bibinfo{year}{2018}).

\bibitem[{\citenamefont{Solntsev et~al.}(2021)\citenamefont{Solntsev, Agarwal,
  and Kivshar}}]{SAK21}
\bibinfo{author}{\bibfnamefont{A.~S.} \bibnamefont{Solntsev}},
  \bibinfo{author}{\bibfnamefont{G.~S.} \bibnamefont{Agarwal}},
  \bibnamefont{and} \bibinfo{author}{\bibfnamefont{Y.}~\bibnamefont{Kivshar}},
  \bibinfo{journal}{Nat.\ Photon.} \textbf{\bibinfo{volume}{15}},
  \bibinfo{pages}{327} (\bibinfo{year}{2021}).

\bibitem[{\citenamefont{Kwiat et~al.}(1995)\citenamefont{Kwiat, Mattle,
  Weinfurter, Zeilinger, Sergienko, and Shih}}]{KMW95}
\bibinfo{author}{\bibfnamefont{P.~G.} \bibnamefont{Kwiat}},
  \bibinfo{author}{\bibfnamefont{K.}~\bibnamefont{Mattle}},
  \bibinfo{author}{\bibfnamefont{H.}~\bibnamefont{Weinfurter}},
  \bibinfo{author}{\bibfnamefont{A.}~\bibnamefont{Zeilinger}},
  \bibinfo{author}{\bibfnamefont{A.~V.} \bibnamefont{Sergienko}},
  \bibnamefont{and} \bibinfo{author}{\bibfnamefont{Y.}~\bibnamefont{Shih}},
  \bibinfo{journal}{Phys.\ Rev.\ Lett.} \textbf{\bibinfo{volume}{75}},
  \bibinfo{pages}{4337} (\bibinfo{year}{1995}).

\bibitem[{\citenamefont{Arnaut and Barbosa}(2000)}]{AB00}
\bibinfo{author}{\bibfnamefont{H.~H.} \bibnamefont{Arnaut}} \bibnamefont{and}
  \bibinfo{author}{\bibfnamefont{G.~A.} \bibnamefont{Barbosa}},
  \bibinfo{journal}{Phys.\ Rev.\ Lett.} \textbf{\bibinfo{volume}{85}},
  \bibinfo{pages}{286} (\bibinfo{year}{2000}).

\bibitem[{\citenamefont{Boyd}(2008)}]{B08_3}
\bibinfo{author}{\bibfnamefont{R.~W.} \bibnamefont{Boyd}},
  \emph{\bibinfo{title}{Nonlinear Optics}} (\bibinfo{publisher}{Academic
  Press}, \bibinfo{address}{Amsterdam}, \bibinfo{year}{2008}),
  \bibinfo{edition}{3rd} ed.

\bibitem[{\citenamefont{van~der Meer et~al.}(2020)\citenamefont{van~der Meer,
  Renema, Brecht, Silberhorn, and Pinkse}}]{VRB20}
\bibinfo{author}{\bibfnamefont{R.}~\bibnamefont{van~der Meer}},
  \bibinfo{author}{\bibfnamefont{J.~J.} \bibnamefont{Renema}},
  \bibinfo{author}{\bibfnamefont{B.}~\bibnamefont{Brecht}},
  \bibinfo{author}{\bibfnamefont{C.}~\bibnamefont{Silberhorn}},
  \bibnamefont{and} \bibinfo{author}{\bibfnamefont{P.~W.}
  \bibnamefont{Pinkse}}, \bibinfo{journal}{Phys.\ Rev.\ A}
  \textbf{\bibinfo{volume}{101}}, \bibinfo{pages}{063821}
  (\bibinfo{year}{2020}).

\bibitem[{\citenamefont{Yang et~al.}(2008)\citenamefont{Yang, Liscidini, and
  Sipe}}]{YLS08}
\bibinfo{author}{\bibfnamefont{Z.}~\bibnamefont{Yang}},
  \bibinfo{author}{\bibfnamefont{M.}~\bibnamefont{Liscidini}},
  \bibnamefont{and} \bibinfo{author}{\bibfnamefont{J.~E.} \bibnamefont{Sipe}},
  \bibinfo{journal}{Phys.\ Rev.\ A} \textbf{\bibinfo{volume}{77}},
  \bibinfo{pages}{033808} (\bibinfo{year}{2008}).

\bibitem[{\citenamefont{Helt et~al.}(2015)\citenamefont{Helt, Steel, and
  Sipe}}]{HSS15}
\bibinfo{author}{\bibfnamefont{L.~G.} \bibnamefont{Helt}},
  \bibinfo{author}{\bibfnamefont{M.~J.} \bibnamefont{Steel}}, \bibnamefont{and}
  \bibinfo{author}{\bibfnamefont{J.~E.} \bibnamefont{Sipe}},
  \bibinfo{journal}{New\ J.\ Phys.} \textbf{\bibinfo{volume}{17}},
  \bibinfo{pages}{013055} (\bibinfo{year}{2015}).

\bibitem[{\citenamefont{Pomarico et~al.}(2009)\citenamefont{Pomarico,
  Sanguinetti, Gisin, Thew, Zbinden, Schreiber, Thomas, and Sohler}}]{PSB09}
\bibinfo{author}{\bibfnamefont{E.}~\bibnamefont{Pomarico}},
  \bibinfo{author}{\bibfnamefont{B.}~\bibnamefont{Sanguinetti}},
  \bibinfo{author}{\bibfnamefont{N.}~\bibnamefont{Gisin}},
  \bibinfo{author}{\bibfnamefont{R.}~\bibnamefont{Thew}},
  \bibinfo{author}{\bibfnamefont{H.}~\bibnamefont{Zbinden}},
  \bibinfo{author}{\bibfnamefont{G.}~\bibnamefont{Schreiber}},
  \bibinfo{author}{\bibfnamefont{A.}~\bibnamefont{Thomas}}, \bibnamefont{and}
  \bibinfo{author}{\bibfnamefont{W.}~\bibnamefont{Sohler}},
  \bibinfo{journal}{New\ J.\ Phys.} \textbf{\bibinfo{volume}{11}},
  \bibinfo{pages}{113042} (\bibinfo{year}{2009}).

\bibitem[{\citenamefont{Luo et~al.}(2015)\citenamefont{Luo, Herrmann, Krapick,
  Brecht, Ricken, Quiring, Suche, Sohler, and Silberhorn}}]{LHK15}
\bibinfo{author}{\bibfnamefont{K.-H.} \bibnamefont{Luo}},
  \bibinfo{author}{\bibfnamefont{H.}~\bibnamefont{Herrmann}},
  \bibinfo{author}{\bibfnamefont{S.}~\bibnamefont{Krapick}},
  \bibinfo{author}{\bibfnamefont{B.}~\bibnamefont{Brecht}},
  \bibinfo{author}{\bibfnamefont{R.}~\bibnamefont{Ricken}},
  \bibinfo{author}{\bibfnamefont{V.}~\bibnamefont{Quiring}},
  \bibinfo{author}{\bibfnamefont{H.}~\bibnamefont{Suche}},
  \bibinfo{author}{\bibfnamefont{W.}~\bibnamefont{Sohler}}, \bibnamefont{and}
  \bibinfo{author}{\bibfnamefont{C.}~\bibnamefont{Silberhorn}},
  \bibinfo{journal}{New\ J.\ Phys.} \textbf{\bibinfo{volume}{17}},
  \bibinfo{pages}{073039} (\bibinfo{year}{2015}).

\bibitem[{\citenamefont{Ilchenko et~al.}(2004)\citenamefont{Ilchenko,
  Savchenkov, Matsko, and Maleki}}]{ISM04}
\bibinfo{author}{\bibfnamefont{V.~S.} \bibnamefont{Ilchenko}},
  \bibinfo{author}{\bibfnamefont{A.~A.} \bibnamefont{Savchenkov}},
  \bibinfo{author}{\bibfnamefont{A.~B.} \bibnamefont{Matsko}},
  \bibnamefont{and} \bibinfo{author}{\bibfnamefont{L.}~\bibnamefont{Maleki}},
  \bibinfo{journal}{Phys.\ Rev.\ Lett.} \textbf{\bibinfo{volume}{92}},
  \bibinfo{pages}{043903} (\bibinfo{year}{2004}).

\bibitem[{\citenamefont{Guo et~al.}(2017)\citenamefont{Guo, ling Zou, Schuck,
  Jung, Cheng, and Tang}}]{GZS17}
\bibinfo{author}{\bibfnamefont{X.}~\bibnamefont{Guo}},
  \bibinfo{author}{\bibfnamefont{C.}~\bibnamefont{ling Zou}},
  \bibinfo{author}{\bibfnamefont{C.}~\bibnamefont{Schuck}},
  \bibinfo{author}{\bibfnamefont{H.}~\bibnamefont{Jung}},
  \bibinfo{author}{\bibfnamefont{R.}~\bibnamefont{Cheng}}, \bibnamefont{and}
  \bibinfo{author}{\bibfnamefont{H.~X.} \bibnamefont{Tang}},
  \bibinfo{journal}{Light\ Sci.\ Appl.} \textbf{\bibinfo{volume}{6}},
  \bibinfo{pages}{e16249} (\bibinfo{year}{2017}).

\bibitem[{\citenamefont{Sun et~al.}(2021)\citenamefont{Sun, Basov, and
  Fogler}}]{SBF21}
\bibinfo{author}{\bibfnamefont{Z.}~\bibnamefont{Sun}},
  \bibinfo{author}{\bibfnamefont{D.}~\bibnamefont{Basov}}, \bibnamefont{and}
  \bibinfo{author}{\bibfnamefont{M.}~\bibnamefont{Fogler}},
  \bibinfo{journal}{arXiv preprint arXiv:2110.14917}  (\bibinfo{year}{2021}).

\bibitem[{\citenamefont{Novotny and Hecht}(2006)}]{NH06}
\bibinfo{author}{\bibfnamefont{L.}~\bibnamefont{Novotny}} \bibnamefont{and}
  \bibinfo{author}{\bibfnamefont{B.}~\bibnamefont{Hecht}},
  \emph{\bibinfo{title}{Principles of Nano-optics}}
  (\bibinfo{publisher}{Cambridge University Press}, \bibinfo{address}{New
  York}, \bibinfo{year}{2006}).

\bibitem[{\citenamefont{De~Rossi and Berger}(2002)}]{DB02}
\bibinfo{author}{\bibfnamefont{A.}~\bibnamefont{De~Rossi}} \bibnamefont{and}
  \bibinfo{author}{\bibfnamefont{V.}~\bibnamefont{Berger}},
  \bibinfo{journal}{Phys.\ Rev.\ Lett.} \textbf{\bibinfo{volume}{88}},
  \bibinfo{pages}{043901} (\bibinfo{year}{2002}).

\bibitem[{\citenamefont{Booth et~al.}(2002)\citenamefont{Booth, Atat{\"u}re,
  Di~Giuseppe, Saleh, Sergienko, and Teich}}]{BAD02}
\bibinfo{author}{\bibfnamefont{M.~C.} \bibnamefont{Booth}},
  \bibinfo{author}{\bibfnamefont{M.}~\bibnamefont{Atat{\"u}re}},
  \bibinfo{author}{\bibfnamefont{G.}~\bibnamefont{Di~Giuseppe}},
  \bibinfo{author}{\bibfnamefont{B.~E.} \bibnamefont{Saleh}},
  \bibinfo{author}{\bibfnamefont{A.~V.} \bibnamefont{Sergienko}},
  \bibnamefont{and} \bibinfo{author}{\bibfnamefont{M.~C.} \bibnamefont{Teich}},
  \bibinfo{journal}{Phys.\ Rev.\ A} \textbf{\bibinfo{volume}{66}},
  \bibinfo{pages}{023815} (\bibinfo{year}{2002}).

\bibitem[{\citenamefont{Orieux et~al.}(2011)\citenamefont{Orieux, Caillet,
  Lema{\^\i}tre, Filloux, Favero, Leo, and Ducci}}]{OCL11}
\bibinfo{author}{\bibfnamefont{A.}~\bibnamefont{Orieux}},
  \bibinfo{author}{\bibfnamefont{X.}~\bibnamefont{Caillet}},
  \bibinfo{author}{\bibfnamefont{A.}~\bibnamefont{Lema{\^\i}tre}},
  \bibinfo{author}{\bibfnamefont{P.}~\bibnamefont{Filloux}},
  \bibinfo{author}{\bibfnamefont{I.}~\bibnamefont{Favero}},
  \bibinfo{author}{\bibfnamefont{G.}~\bibnamefont{Leo}}, \bibnamefont{and}
  \bibinfo{author}{\bibfnamefont{S.}~\bibnamefont{Ducci}},
  \bibinfo{journal}{J.\ Opt.\ Soc.\ Am.\ B} \textbf{\bibinfo{volume}{28}},
  \bibinfo{pages}{45} (\bibinfo{year}{2011}).

\bibitem[{\citenamefont{Saravi et~al.}(2017)\citenamefont{Saravi, Pertsch, and
  Setzpfandt}}]{SPS17}
\bibinfo{author}{\bibfnamefont{S.}~\bibnamefont{Saravi}},
  \bibinfo{author}{\bibfnamefont{T.}~\bibnamefont{Pertsch}}, \bibnamefont{and}
  \bibinfo{author}{\bibfnamefont{F.}~\bibnamefont{Setzpfandt}},
  \bibinfo{journal}{Phys.\ Rev.\ Lett.} \textbf{\bibinfo{volume}{118}},
  \bibinfo{pages}{183603} (\bibinfo{year}{2017}).

\bibitem[{\citenamefont{Plenio and Virmani}(2014)}]{P14}
\bibinfo{author}{\bibfnamefont{M.~B.} \bibnamefont{Plenio}} \bibnamefont{and}
  \bibinfo{author}{\bibfnamefont{S.~S.} \bibnamefont{Virmani}},
  \emph{\bibinfo{title}{An Introduction to Entanglement Theory}}
  (\bibinfo{publisher}{Springer International Publishing},
  \bibinfo{address}{Cham}, \bibinfo{year}{2014}), pp.
  \bibinfo{pages}{173--209}.

\bibitem[{\citenamefont{Dmitriev et~al.}(1999)\citenamefont{Dmitriev,
  Gurzadyan, and Nikogosyan}}]{DGN99}
\bibinfo{author}{\bibfnamefont{V.~G.} \bibnamefont{Dmitriev}},
  \bibinfo{author}{\bibfnamefont{G.~G.} \bibnamefont{Gurzadyan}},
  \bibnamefont{and} \bibinfo{author}{\bibfnamefont{D.~N.}
  \bibnamefont{Nikogosyan}}, \emph{\bibinfo{title}{Handbook of Nonlinear
  Optical Crystals}}, vol.~\bibinfo{volume}{64}
  (\bibinfo{publisher}{Springer-Verlag}, \bibinfo{address}{Berlin},
  \bibinfo{year}{1999}), \bibinfo{edition}{3rd} ed.

\bibitem[{\citenamefont{{Garc\'{\i}a de Abajo}
  et~al.}(2003)\citenamefont{{Garc\'{\i}a de Abajo}, Rivacoba, Zabala, and
  Echenique}}]{paper047}
\bibinfo{author}{\bibfnamefont{F.~J.} \bibnamefont{{Garc\'{\i}a de Abajo}}},
  \bibinfo{author}{\bibfnamefont{A.}~\bibnamefont{Rivacoba}},
  \bibinfo{author}{\bibfnamefont{N.}~\bibnamefont{Zabala}}, \bibnamefont{and}
  \bibinfo{author}{\bibfnamefont{P.~M.} \bibnamefont{Echenique}},
  \bibinfo{journal}{Phys.\ Rev.\ B} \textbf{\bibinfo{volume}{68}},
  \bibinfo{pages}{205105} (\bibinfo{year}{2003}).

\bibitem[{\citenamefont{Agrawal}(2000)}]{A00}
\bibinfo{author}{\bibfnamefont{G.~P.} \bibnamefont{Agrawal}}, in
  \emph{\bibinfo{booktitle}{Nonlinear Science at the Dawn of the 21st Century}}
  (\bibinfo{publisher}{Springer}, \bibinfo{address}{Berlin},
  \bibinfo{year}{2000}).

\bibitem[{\citenamefont{Keiser}(2011)}]{K11}
\bibinfo{author}{\bibfnamefont{G.}~\bibnamefont{Keiser}},
  \emph{\bibinfo{title}{Optical Fiber Communications}}
  (\bibinfo{publisher}{McGraw-Hill}, \bibinfo{year}{2011}).

\bibitem[{\citenamefont{Marcuse}(2013)}]{M13_1}
\bibinfo{author}{\bibfnamefont{D.}~\bibnamefont{Marcuse}},
  \emph{\bibinfo{title}{Theory of Dielectric Optical Waveguides}}
  (\bibinfo{publisher}{Elsevier}, \bibinfo{year}{2013}).

\bibitem[{\citenamefont{Engelbrecht}(2015)}]{B15}
\bibinfo{author}{\bibfnamefont{R.}~\bibnamefont{Engelbrecht}},
  \emph{\bibinfo{title}{Nichtlineare Faseroptik: Grundlagen Und
  Anwendungsbeispiele}} (\bibinfo{publisher}{Springer-Verlag},
  \bibinfo{address}{Belin}, \bibinfo{year}{2015}).

\bibitem[{\citenamefont{Saleh and Teich}(2019)}]{ST19}
\bibinfo{author}{\bibfnamefont{B.~E.} \bibnamefont{Saleh}} \bibnamefont{and}
  \bibinfo{author}{\bibfnamefont{M.~C.} \bibnamefont{Teich}},
  \emph{\bibinfo{title}{Fundamentals of Photonics}} (\bibinfo{publisher}{John
  Wiley \& Sons}, \bibinfo{address}{New York}, \bibinfo{year}{2019}).

\bibitem[{\citenamefont{Yeh and Shimabukuro}(2008)}]{YS08_1}
\bibinfo{author}{\bibfnamefont{C.}~\bibnamefont{Yeh}} \bibnamefont{and}
  \bibinfo{author}{\bibfnamefont{F.~I.} \bibnamefont{Shimabukuro}},
  \emph{\bibinfo{title}{The Essence of Dielectric Waveguides}}
  (\bibinfo{publisher}{Springer}, \bibinfo{address}{Berlin},
  \bibinfo{year}{2008}).

\bibitem[{\citenamefont{Abramowitz and Stegun}(1972)}]{AS1972}
\bibinfo{author}{\bibfnamefont{M.}~\bibnamefont{Abramowitz}} \bibnamefont{and}
  \bibinfo{author}{\bibfnamefont{I.~A.} \bibnamefont{Stegun}},
  \emph{\bibinfo{title}{Handbook of Mathematical Functions}}
  (\bibinfo{publisher}{Dover}, \bibinfo{address}{New York},
  \bibinfo{year}{1972}).

\bibitem[{{\relax DLMF}()}]{DLMF}
{\relax DLMF}, \emph{\bibinfo{title}{{\it nist digital library of mathematical
  functions}}}, \bibinfo{howpublished}{http://dlmf.nist.gov/, Release 1.1.3 of
  2021-09-15}, \bibinfo{note}{f.~W.~J. Olver, A.~B. {Olde Daalhuis}, D.~W.
  Lozier, B.~I. Schneider, R.~F. Boisvert, C.~W. Clark, B.~R. Miller, B.~V.
  Saunders, H.~S. Cohl, and M.~A. McClain, eds.},
  \urlprefix\url{http://dlmf.nist.gov/}.

\end{thebibliography}

\clearpage %--- optional
\pagebreak \onecolumngrid \section*{SUPPLEMENTARY FIGURES} %---SI---arxiv optional

% Figure S1 ------------------------------------------------
\begin{figure*}[h] 
\centering \includegraphics[width=0.75\textwidth]{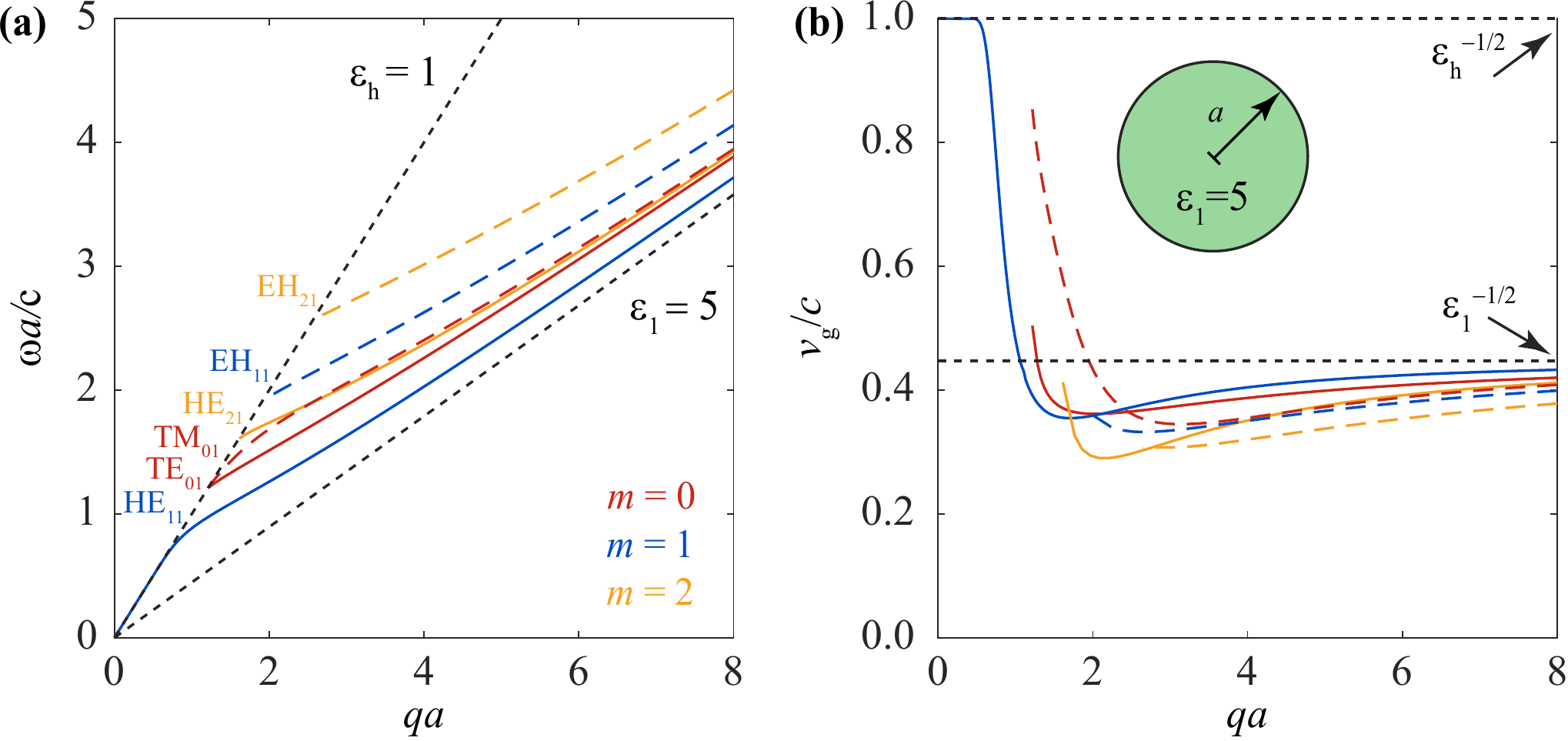}
\caption{{\bf Dispersion relation of guided modes.} {\bf (a)} Dispersion relations of the lowest-orders propagating modes in a cylindrical waveguide of radius $a$ made of $\epsc=5$ material and hosted in air ($\epsh=1$). We consider the lowest-order solutions with azimuthal and radial numbers $m=0-2$ and $l=1$. Mode labels follow the notation TE$_{0l}$ and TM$_{0l}$ for $m=0$ (transverse electric and magnetic modes, respectively), as well as EH$_{ml}$ and HE$_{ml}$ for $m\neq0$ (see Sec.\ \ref{Sec4.1} in the main text). {\bf (b)} Group velocities corresponding to the modes in (a). Black dashed lines in both panels indicate the light cones inside and outside the waveguide.} \label{FigS1}
\end{figure*}

\end{document}